\documentclass[a4paper,11pt]{article}
\usepackage{amsmath,amsfonts,amssymb,times,mathptmx,graphicx,amsthm,lineno}
\usepackage{placeins}
\usepackage[T1]{fontenc}
\usepackage[latin1]{inputenc}

\title{\bf Properties of the Center of Gravity as an Algorithm for
Position Measurements: Two-Dimensional Geometry }
\author{Gregorio Landi$ $\thanks{Corresponding
author. Gregorio.Landi@fi.infn.it}\\
\\
\llap{$ $} Dipartimento di Fisica e Astronomia,
Universita' di Firenze and INFN\\
Largo E. Fermi 2 (Arcetri) 50125, Firenze, Italy\\ \\
}

\date{\em October 5, 2002 and February 26, 2021}

\begin{document}
\maketitle 
\begin{abstract}
The center of gravity as an algorithm for position measurements is
analyzed for a two-dimensional geometry. Several mathematical
consequences of discretization for various types of detector
arrays are extracted. Arrays with rectangular, hexagonal, and
triangular detectors are analytically studied, and tools are given
to simulate their discretization properties. Special signal
distributions free of discretized error are isolated. It is proved
that some crosstalk spreads are able to eliminate the center of
gravity discretization error for any signal distribution (ideal detectors).
Simulations, adapted to the CMS em-calorimeter and to a triangular
detector array, are provided for energy and position
reconstruction algorithms with a finite number of detectors.
\end{abstract}




\tableofcontents

\pagenumbering{arabic} \oddsidemargin 0cm  \evensidemargin 0cm

\section{Introduction 2021}

The aim of this paper was the study of the e.m. calorimeters,
the pixel detectors were in an early stages and we neglected them. 
The availability of data on silicon micro-strip detectors
deviated our attention from this aim, and from further
extension to a three-dimensional spherical geometry.
Preliminary steps in this last direction showed error patterns
very similar to those in the following, with deep valleys
near to exact reconstructions. However, the complexity
of the Bessel j-functions suggested an early abandon of the
project for the probably lack of readers.    
Instead, the silicon micro-strips are an ideal one dimension system,
perfect for ref.~\cite{landi01}. The deep analysis
of those data, evidenced the impossibility of a single
variance for each hit (homoscedasticity), as it is
usually assumed in track fitting. The study of methods to handle
those probabilities absorbed all our attentions and
produced very interesting new results. This shift of
interests, did not allow applications of the two-dimension
COG properties, even if the recent use of the silicon
pixel-detectors would be a perfect combination of tracking
and positioning. However, the difficulty to access to data hinders
this possibility.  The complexity of the developments here
described requires a mathematical skill not frequent among the
experimental staffs. Unfortunately, the reconstruction
algorithms are complex mathematical instruments and any data
analysis can not avoid those complications.
Some printing errors, due to the splitting of the equations in 
two columns, were corrected. The explicit definition of the ideal detector is 
introduced also for this two-dimensional geometry, as we did for the 
one dimension system in arXiv:1908.04447. In successive works, these types 
of detectors were indicated in this synthetic way, sure of its introduction
here and in ref.~\cite{landi01}. Among the beneficial effects of the absence 
of the discretization errors (ideal detectors), it must be recalled its 
effectiveness in the noise regularization with a drastic simplification in 
the handling of the detector heteroscedasticity.

\section{Introduction}

The center of gravity (COG) is a widespread method of pattern
recognition used in experimental works as a tool to improve
position reconstruction from sets of measurements. Our previous
work~\cite{landi01} was devoted to the introduction of methods to
define COG properties when the COG is applied. The
approach~\cite{landi01} was limited to one-dimensional systems,
and only  a marginal attention was devoted to two-dimensional
systems, when they can be reverted to two one-dimensional systems.
This condition covers only a restricted class of two-dimensional
detectors: Array of rectangular detectors with constant
efficiency, no loss at the border, the COG algorithm calculated
with an infinite row of detectors in one direction. Thus, the
dependence of the signal distribution on $y$ does not contribute
to the COG calculation in $x$, and the two-dimensional problem is
split into two one-dimensional configurations. These limitations
could be very unrealistic in a few cases, even for detectors of
the indicated type. Detectors of different shapes, hexagonal,
triangular etc. can never be reduced to two one-dimensional
configurations.

Generalization to a two-dimensional geometry entails additional
complications compared to ~\cite{landi01}, and many more forms and
setups are allowed. In this work, we will limit our considerations
to parallelogram, hexagon, and triangle detector arrays. These
forms cover almost all the types of detectors used in high-energy
physics. Although the triangular forms have not been widely used
in recent years, the Crystal Ball detector ~\cite{crystball} has
an em-calorimeter with triangular-section crystals. The study of
the triangular forms allows us to determine the method to treat
far more complex detector arrays.

As will be evident, this work is largely based on the results of
~\cite{landi01}, but due to the necessity of a two-dimensional
generalization and the special handling required by the hexagons
and triangles, almost all the equations will be derived in the new
context. Array periodicity is a fundamental tool for our
derivations. A group of detectors can easily be approximated with
a periodic lattice owing to the finite---generally small---range
of the signal distributions that affects a small number of
detectors, which can almost always be replicated as a periodic
mosaic. For this reason, Section 2 introduces the initial
definitions and the general equations that are valid for any type
of periodic detector array.

In Section 3, the equations are applied to three types of detector
array: Parallelograms, hexagons, and triangles. The triangle array
exemplifies the way of handling any type of detector mosaic.

Section 4 is devoted to the study of the signal distributions
whose COGs are free of discretization errors. It will be evident
that the parallelograms, hexagons, and triangles are an extremely
reduced sample of the class of signal distributions that have
exact COGs at particular sizes. It is surprising how large this
class is compared to the one-dimensional case.

Section 5 deals with the modifications introduced by loss and
crosstalk. Loss destroys the splitting of a two-dimensional
problem into two one dimensions, even for an array of rectangular
detectors. There is a large set of crosstalk functions that saves
the norm and spreads the incident signal among the detectors in
such a way as to eliminate the COG errors for any type of signal
distribution. We will also determine the connection to other
mathematical properties far from the COG problems.

In Section 6, array periodicity is abandoned and a method to
handle a finite set of detectors is defined. The method is applied
to the CMS and Crystal Ball em-calorimeter.

Section 7 concludes and summarizes the results.
\section{Sampling in Two Dimensions}
\subsection{Definition of the Formalism}

To introduce some notation, let us consider an array of
rectangular detectors as shown in
\begin{figure}[h!]
\begin{center}
\includegraphics[scale=0.8]{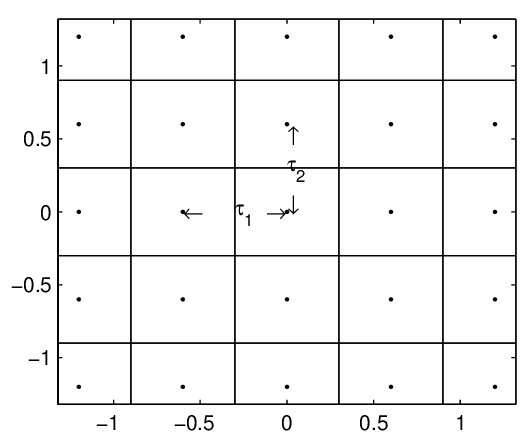}
\end{center}
\caption{\em Array of rectangular detectors, $\tau_1$ is the
$x$-distance of the centers of two neighboring detectors, $\tau_2$
is the $y$-distance. }
 \label{fig:prima}
\end{figure}
Fig.~\ref{fig:prima}. The two constants $\tau_1$ and $\tau_2$
establish the size of the array; $\tau_1$ is the distance from the
center of a detector to the center of its neighbor in
$x$-direction, $\tau_2$ is the same in $y$-direction. If
$\tau_1=\tau_2$, the detectors have a square section.

As in ~\cite{landi01}, we will mainly consider average signal
distributions on which we impose some symmetry conditions deriving
from the assumed homogeneity of the medium where the signal
distribution is produced. These conditions are not essential to
our procedures. The equations are able to handle nonsymmetric
signal distributions as well, but the asymmetry is rarely
experimentally measurable, thereby forcing us to consider averages
over asymmetries. Simulation differs in that the asymmetric signal
distributions can be easily generated and accounted for. Other
implicit conditions are positivity and connectedness of the signal
distribution (an em shower in a calorimeter); no limitations of
these type will affect the equations, and we will give examples of
nonconnected signal distributions.

Let us define $\varphi(\mathbf{x})$ as the signal distribution. As
vectorial notation will be used almost everywhere, $\mathbf{x}$
indicates variables $x,y$. We will abandon the vectorial notation
when the expressions are not overly long. On
$\varphi(\mathbf{x})$, we impose the following two conditions:
\begin{align}
&\int_{\mathbb{R}^{2}}\mathrm{d}\mathbf{x}\
\mathbf{x}\varphi(\mathbf{x})=0,\label{eq:uno}\\
&\int_{\mathbb{R}^{2}}\ \mathrm{d}\mathbf{x}\
\varphi(\mathbf{x})=1.\label{eq:tre}
\end{align}
Equation~(\ref{eq:uno}) fixes the position of the two COGs
$\mathbf{r}_g=(x_g,y_g)$ of $\varphi(\mathbf{x})$ at the origin of
the axes, where as explained in Fig.~\ref{fig:prima}, the COG of a
detector is located. It is clear that a symmetric signal
distribution $\varphi(\mathbf{x})=\varphi(-\mathbf{x})$ satisfies
the two equations. Equation~(\ref{eq:tre}) is a normalization
condition that eliminates a constant in some of the following
procedures.

A good detector generally performs a linear transformation on the
collected signal. For example, a good crystal of an em-calorimeter
integrates the energy converging on it. The most general linear
transformation on the signal distribution induced by a detector is
the convolution:
\begin{equation}
\mathrm{f}(\mathbf{x})=\int_{\mathbb{R}^{2}}\mathrm{g}(\mathbf{x}-\mathbf{x}')
\varphi(\mathbf{x}')\,\mathrm{d}\mathbf{x}\ ,
 \label{eq:conv}
\end{equation}
where g$(\mathbf{x})$ embodies the effects of the detector. If
$\varphi(\mathbf{x})$ is shifted from the origin by a vector
$\boldsymbol\varepsilon=(\varepsilon_1;\varepsilon_2)$ in
$x$-direction and in $y$-direction, f$(\mathbf{x})$ becomes:
\begin{equation}
{\mathrm{f}_{\boldsymbol\varepsilon}}(\mathbf{x})=\int_{\mathbb{R}^{2}}\mathrm{g}(\mathbf{x}-\mathbf{x}')
\varphi(\mathbf{x}'-\boldsymbol\varepsilon)\,\mathrm{d}\mathbf{x}'\ .
 \label{eq:conv1}
\end{equation}
For the properties of the convolution~\cite{libroFT} we get:
\begin{equation*}
\mathrm{f}_{\boldsymbol\varepsilon}(\mathbf{x})=\int_{\mathbb{R}^{2}}
\mathrm{g}(\mathbf{x}-\boldsymbol\varepsilon-\mathbf{x}')
\varphi(\mathbf{x}')\,\mathrm{d}\mathbf{x}'\ ,
\end{equation*}
so f$_{\boldsymbol\varepsilon}(\mathbf{x})$ becomes:
\begin{equation}
\mathrm{f}_{\boldsymbol\varepsilon}(\mathbf{x})=\mathrm{f}(\mathbf{x}-\boldsymbol\varepsilon)\
. \label{eq:sei}
\end{equation}
With the normalization of Eq.~(\ref{eq:tre}) and assuming
Eq.~(\ref{eq:uno}) even for g$(\mathbf{x})$, the definition of the
COG is proportional to the first moment of
f$_{\boldsymbol\varepsilon}(\mathbf{x})$. Convolution
theorems~\cite{libroFT} state that the first moment of
f$_{\boldsymbol\varepsilon}(\mathbf{x})$ is the sum of the first
moment of g$(\mathbf{x})$ and the first moment of
$\varphi(\mathbf{x}-\boldsymbol\varepsilon)$. Hence, for our
choice of the reference system, the COG of
f$_{\boldsymbol\varepsilon}(\mathbf{x})$ is
$\boldsymbol\varepsilon$. At this level, the linear transformation
of Eq.~(\ref{eq:conv1}) on the signal distribution does not modify
the COGs positions.

\subsection{Sampling on a Rectangular Lattice}

It is evident that Eq.~(\ref{eq:conv1}) is not the true
transformation performed by the detector on the signal. The
function f$_{\boldsymbol\varepsilon}(\mathbf{x})$ is not measured
for any $\mathbf{x}\in\mathbb{R}^2$ nor does the detector
continuously "scan" the signal distribution. The measuring device
is a set of detectors that are well fixed in definite spatial
points, and they give only a sampling of the function
f$_{\boldsymbol\varepsilon}(\mathbf{x})$ with very few points
above the detector noise. This discretization is the principal
source of systematic error of the COG as a position reconstruction
algorithm.

The detectors in the array will be considered identical and
arranged in a periodic rectangular lattice. To simplify the
notations, we shall consider in the following
$n,l\in\mathbb{Z}^2$, and in absence of other indication our sums
will run over $n,l\in\mathbb{Z}^2$. We will indicate the constants
of the sampling lattice as $ \mathbf{d}_1=(\tau_1;0)$ and $
\mathbf{d}_2=(0;\tau_2)$. If
$\big\{$f$_{\boldsymbol\varepsilon}(n\mathbf{d}_{1}+l\mathbf{d}_{2})\big\}$
is a set of sampled values extracted from
f$_{\boldsymbol\varepsilon}(\mathbf{x})$,  the COG
$\mathbf{r}_g=(x_g;y_g)$ of the set is given by (with $x_g$ the
COG in $x$-direction and $y_g$ the COG in $y$-direction):
\begin{equation}
\mathbf{r}_{g}=\sum_{n,l}\
(n\mathbf{d}_{1}+l\mathbf{d}_2)\mathrm{f}_{\boldsymbol\varepsilon}(n\mathbf{d}_{1}+l\mathbf{d}_{2})
\ \ \ \ \
\sum_{n,l}\mathrm{f}_{\boldsymbol\varepsilon}(n\mathbf{d}_{1}+l\mathbf{d}_{2})=1\
. \label{eq:cog1}
\end{equation}
Assuming a lossless detector array, the sampling of
f$_{\boldsymbol\varepsilon}(\mathbf{x})$ saves the normalization.
In the presence of loss, the normalization is no longer saved, and
$\mathbf{r}_g$ can be recast in the form:
\begin{equation}
\mathbf{r}_{g}(\boldsymbol\varepsilon)=\boldsymbol\varepsilon+
\frac{\sum_{n,l}(n\mathbf{d}_{1}+l\mathbf{d}_{2}-
\boldsymbol\varepsilon)\mathrm{f}(n\mathbf{d}_{1}+l\mathbf{d}_{2}-\boldsymbol\varepsilon)}
{\sum_{n,l}\mathrm{f}(n\mathbf{d}_{1}+l\mathbf{d}_{2}-\boldsymbol\varepsilon)}
\label{eq:cog2} ,
\end{equation}
where we use Eq.~(\ref{eq:sei}) and the difference of
$\mathbf{r}_g$ (the COG position) with respect to the exact
position $\boldsymbol\varepsilon$ is isolated. This vectorial
notation seems to mix rows of detectors (index $n$) and columns
(index $l$) in the equations for $\mathbf{r}_g$. The two vectors
$\mathbf{d}_1$ and $\mathbf{d}_2$ are parallel respectively to the
$x$ and $y$ axes, and their orthogonality eliminates the mixture.
In the following, the nonorthogonal lattices will explicitly
require such mixture.

The modifications of Eq.~(\ref{eq:cog2}) to deal with a subset of
detectors are straightforward. The sum on $n$ and $l$ must cover
only the detectors considered. More developments are required to
highlight the dependence on the system parameters, i.e., detector
efficiency and shape and signal distribution and position.

Let us prove further properties of Eqs.~(\ref{eq:cog1}).
In~\cite{landi01}, we studied a  general efficiency function
g$(x)$ that could be different from zero even for values of $x$
outside the space of a single detector (crosstalk). We called
uniform crosstalk the efficiency functions g$(x)$ having
{\em{almost everywhere}} (a. e.) the property
$\Sigma_{n}$g$(n\tau-x)=1$. The definition is extended for a
two-dimensional system (here and in the following g$(\mathbf{x})$
is normalized as a detector of unitary efficiency and area
$\tau_1\tau_2$) :
\begin{equation}
\sum_{n,l}\mathrm{g}(n\mathbf{d}_{1}+l\mathbf{d}_{2}-\mathbf{x})=1\
\ \ \ \mathit{(a. e.).} \label{eq:uniform}
\end{equation}
The condition of validity for Eq. (\ref{eq:uniform}) a. e. means
that in a null set (for example, at the detector boundaries)
violations are allowed. These violations have no effect on
Eq.~(\ref{eq:cog1}) due to the convolution operation on the
function g$(\mathbf{x})$ that eliminates all them. The notation of
Eq.~(\ref{eq:uniform}) generalizes the intuition of an infinite
array of identical detectors each with unitary efficiency packed
without holes and area $\tau_1\tau_2$. Evidently,
Eq.~(\ref{eq:uniform}) deals with more complex configurations than
this naive intuition: It automatically contains the crosstalk that
distributes part of the signal collected by a detector to nearby
ones. It is easy to prove the normalization conservation in the
presence of uniform crosstalk:
\begin{equation*}
\sum_{n,l}\mathrm{f}(n\mathbf{d}_{1}+l\mathbf{d}_{2}-\boldsymbol\varepsilon)=
\int_{\mathbb{R}^{2}}\big[\sum_{n,l}\mathrm{g}
(n\mathbf{d}_{1}+l\mathbf{d}_{2}-\boldsymbol{\varepsilon}-
\mathbf{x}')\big]\varphi(\mathbf{x}')\,\mathrm{d}\mathbf{x}'\
,
\end{equation*}
and, applying Eq.~(\ref{eq:uniform}) in parentheses, the
right-hand side becomes the normalization of
$\varphi(\mathbf{x})$. The condition for the absence of
discretization error in $\mathbf{r}_g$ can be extracted similarly
from Eq.~(\ref{eq:cog2}):
\begin{align*}
&\mathbf{r}_{g}=\pmb{\varepsilon+}\sum_{n,l}
(n\mathbf{d}_{1}+l\mathbf{d}_2-\pmb{\varepsilon})\mathrm{f}(n\mathbf{d}_{1}+l\mathbf{d}_{2}-
\pmb\varepsilon)\\
&\mathbf{r}_{g}=\pmb{\varepsilon}+\int_{\mathbb{R}^{2}}\big[\sum_{n,l}
(n\mathbf{d}_{1}+l\mathbf{d}_{2}-\pmb{\varepsilon}-\mathbf{x}')\
\mathrm{g}(n\mathbf{d}_{1}+l\mathbf{d}_{2}-\pmb{\varepsilon}-\mathbf{x}')\big]
\varphi(\mathbf{x}')\,\mathrm{d}\mathbf{x}'\ .
\end{align*}
The $-\mathbf{x}'$ term introduced in the parentheses gives no
contribution to the integral due to
Eqs.~(\ref{eq:uniform},\ref{eq:uno}). Therefore, the condition for
the absence of discretization error in $\mathbf{r}_g$ {\em for any
signal distribution} is given by:
\begin{equation}
\sum_{n,l}(n\mathbf{d}_{1}+l\mathbf{d}_{2}-\mathbf{x})\
\mathrm{g}(n\mathbf{d}_{1}+l\mathbf{d}_{2}-\mathbf{x})=0\ \ \
\mathit{(a. e.)}\ . \label{eq:nocog}
\end{equation}
 In the following, we will show how to construct g-functions with
the property of Eq.~(\ref{eq:nocog}).

\subsection{Sampling on a Parallelogram Lattice}

To handle more complex, albeit periodic, detector arrays, we have
to abandon the simple symmetry of the rectangular array of
sampling points and move
\begin{figure}[h!]
\begin{center}
\includegraphics[scale=0.7]{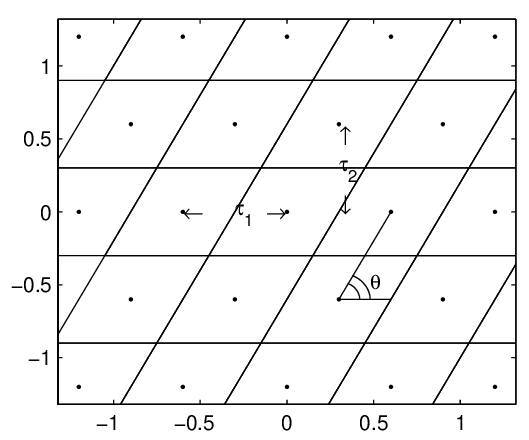}
\end{center}
\caption{\em Array of parallelogram detectors: $\tau_1$ is the
$x$-distance of the centers of two neighboring detectors, $\tau_2$
is the $y$-distance, $\theta$ is the bending angle of the inclined
sides. }
 \label{fig:due}
\end{figure}
on to a more general set of samples organized as an array of
parallelograms (anomalous torus), as given in Fig.~\ref{fig:due}.
On $x$, the family of sampling points is given by horizontal
straight-lines; on $y$, the family of sampling points is given by
the straight-lines at an angle $\vartheta$ with the horizontal.
Our reference system is oriented to have this setup; if rotations
are present, these are charged to the signal distributions. As
above, $\tau_1$ is the distance of the two consecutive detector
centers in direction $x$, and $\tau_2$ is the distance of two
consecutive detector centers in direction $y$,
$\alpha=\tan(\theta)$ is the angular coefficient of the inclined
lines. At first glance, this array of sampling points suggests
parallelogram detector forms, where the rectangular array is given
by $\alpha\rightarrow\infty$. In reality, many detector forms are
allowed, and the arrays of hexagonal and triangular detectors can
be sampled as in Fig.~\ref{fig:due}. This type of symmetry is
connected to the periodic plane tessellation.

We will not even attempt to list all the detector arrays which are
covered by a sampling along the vertices of a parallelogram. For
example, all plane tessellation images by Escher~\cite{Escher} are
created by pictures that can be sampled along two families of
parallel straight lines, whereas the Penrose tiles~\cite{Penrose}
are not. Fortunately, no use of these forms is forecast in real
detectors. Theoretically, the equations we are developing can
handle the exotic forms of Ref.~\cite{Escher} and a subset of the
Penrose tiles.

Neglecting the detector form complications on sampling schema, the
linear effect of a single detector on $\varphi(\mathbf{x})$ is
always contained in Eq.~(\ref{eq:conv}). The sampled version of
Eq.~(\ref{eq:conv}) has a few differences. Formally, the sampling
is generated by multiplying continuous functions with sums of
Dirac $\delta$-functions that record the sampling
points~\cite{libroFT2}. In a periodic array, the $\alpha$ value is
not univocally defined. We can define $\alpha$ with any detector
in the upper line (Fig.~\ref{fig:due}), then
$\alpha_{\rho}^{-1}=\alpha_{0}^{-1}+\rho\tau_1/\tau_2$ where
$\rho\in \mathbb{Z}$, and $\alpha_{0}^{-1}$ is the minimum of the
positive values of $\alpha_{\rho}^{-1}$ . Any other
$\alpha_{\rho}$ simply relabels the detectors. To avoid
ambiguities, we will always use $\alpha=\alpha_0$. Now the lattice
constants are $\mathbf{n}_1=(\tau_1,\ 0)$ and
$\mathbf{n}_2=(\tau_2/\alpha,\ \tau_2)$. The vector $\mathbf{n}_2$
is no longer parallel to the $y$-axis and introduces a mixture of
$n$ and $l$ sums. The vectorial notation hides this difference
with respect to the orthogonal sampling, but it is fundamental in
the applications.

If s$_{\pmb{\varepsilon}}(\mathbf{x})$ is the result of the
sampling operating on f$_{\pmb{\varepsilon}}(\mathbf{x})$, it
can be expressed as:
\begin{equation}\label{eq:sampling}
\mathrm{s}_{\pmb\varepsilon}(\mathbf{x})=
\sum_{n,l}\delta(\mathbf{x}-n\mathbf{n}_{1}-l\mathbf{n}_2)\
\mathrm{f}(n\mathbf{n}_{1}+l\mathbf{n}_{2}-\pmb\varepsilon),
\end{equation}
its Fourier Transform (FT) is defined as usual
($\pmb{\omega}=(\omega_x,\ \omega_y)$):
\begin{equation*}
\mathrm{S}_{\pmb{\varepsilon}}(\pmb{\omega})=
\int_{\mathbb{R}^{2}}\mathrm{e}^{-i{\pmb{\omega}}\cdot\mathbf{x}}
\mathrm{s}_{\pmb{\varepsilon}}(\mathbf{x})\ \mathrm{d}\mathbf{x}
\end{equation*}
or, using Eq.~(\ref{eq:sampling}),
$\mathrm{S}_{\pmb\varepsilon}(\pmb\omega)$
 becomes:
\begin{equation}
\mathrm{S}_{\pmb\varepsilon}(\pmb\omega)
=\sum_{n,l}\mathrm{e}^{-i\pmb\omega\cdot(n\mathbf{n}_{1}+l\mathbf{n}_{2})}\:
\mathrm{f}\,(n\mathbf{n}_{1}+l\mathbf{n}_{2}-\pmb\varepsilon).
\label{eq:samplFT}
\end{equation}
Now $\mathbf{r}_g$ can be defined through
$\mathrm{S}_{\pmb\varepsilon}(\pmb\omega)$ with:
\begin{equation}
\mathbf{r}_{g}=\frac{i}{\mathrm{S}_{\pmb\varepsilon}(0)}
\pmb\nabla_{\pmb\omega}\,\mathrm{S}_{\pmb\varepsilon}
(\pmb\omega) \Big|_{\pmb\omega\rightarrow 0}\ .
\label{eq:FTcog}
\end{equation}
If the normalization of $\varphi(\mathbf{x})$ is conserved (i.e.,
all the energy released in a calorimeter is collected),
$\mathrm{S}_{\pmb\varepsilon}(0)$ is equal to one, and
Eq.~(\ref{eq:samplFT}) yields:
\begin{equation}\label{eq:normS}
\mathrm{S}_{\pmb\varepsilon}(0)=\sum_{n,l}
\mathrm{f}(n\mathbf{n}_{1}+l\mathbf{n}_{2}-\pmb\varepsilon)=1\
\ \ \ \forall\,\pmb\varepsilon\,.
\end{equation}
Equation (\ref{eq:normS}) is an evident generalization of a
similar equation for the rectangular array. The application of
Eq.~(\ref{eq:FTcog}) to the form (\ref{eq:samplFT}) of
$\mathrm{S}_{\pmb\varepsilon}(\pmb\omega)$
 gives a trivial generalization of the
COG expressions for this kind of detector array, but does not
produce any further results. On the contrary, going through the FT
of f$(\mathbf{x})$ and summing over the infinite set of sampling
points, more workable expressions can be obtained. The unusual
form of the sampling steps requires that we give a few details of
the derivation temporarily without vectorial notation. With its
FT,
f$(n\tau_{1}+\frac{l\tau_{2}}{\alpha}-\varepsilon_{1},l\tau_{2}-\varepsilon_{2})$
can be expressed by:
\begin{equation*}
\mathrm{f}(n\tau_{1}+\frac{l\tau_{2}}{\alpha}-\varepsilon_{1},l\tau_{2}-\varepsilon_{2})=
\int_{\mathbb{R}^{2}}\frac{\mathrm{d}\omega_{x}}{2\pi}
\frac{\mathrm{d}\omega_{y}}{2\pi}\,
\mathrm{e}^{i\omega_{x}(n\tau_{1}+\frac{l\tau_{2}}{\alpha}-\varepsilon_{1})}\,
\mathrm{e}^{i \omega_{y}(l\tau_{2}-\varepsilon_{2})}\,\mathrm{F}
(\omega_{x} , \omega_{y})\,  {,}
\end{equation*}
where F$(\omega_{x},\omega_{y})$ is the FT of f$(x,y)$.
Substituting this expression in (\ref{eq:samplFT}), and performing
the sums over the indices $n$ and $l$ with the following formal
relation:
\begin{equation*}
\sum_{k=-\infty}^{+\infty} \mathrm{e}^{-i(\omega-\omega ')k\tau}=
\frac{2\pi}{\tau}\sum_{m=-\infty}^{+\infty} \delta (\omega
-\omega' - \frac{2\pi}{\tau}m)
\end{equation*}
we obtain a special type of Poisson identity~\cite{libroFT2}
adapted to this unusual sampling scheme:
\begin{equation}\label{eq:poisson}
\begin{aligned}
\mathrm{S}_{\varepsilon_{1},\varepsilon_{2}}(\omega_{x},\omega_{y})=&\frac{1}{\tau_{1}\tau_{2}}
\sum_{m,k=-\infty}^{+\infty}\mathrm{e}^{-i(\omega_{x}-\frac{2m\pi}{\tau_{1}})\varepsilon_{1}}\,
\mathrm{e}^{-i(\omega_{y}+\frac{2m\pi}{\tau_{1}\alpha}-\frac{2k\pi}{\tau_{2}})\varepsilon_{2}}
\\
&\mathrm{F}(\omega_{x}-\frac{2m\pi}{\tau_{1}},\omega_{y}+\frac{2m\pi}{\tau_{1}\alpha}-
\frac{2k\pi}{\tau_{2}}),\\
\mathrm{S}_{\pmb\varepsilon}(\pmb\omega)=&\frac{1}{\tau_{1}\tau_{2}}
\sum_{m,k}\mathrm{e}^{-i(\pmb\omega-m\mathbf{r}_{1}-k\mathbf{r}_{2})\cdot\pmb\varepsilon}\,
\mathrm{F}(\pmb\omega-m\mathbf{r}_{1}-k\mathbf{r}_{2}),\\
\mathbf{r}_{1}=&\big(\frac{2\pi}{\tau_{1}}\,;\,-\frac{2\pi}{\alpha\tau_{1}}\big)\\
\mathbf{r}_{2}=&\big(0\ ;\,\frac{2\pi}{\tau_2}\big)
\end{aligned}
\end{equation}
where $\mathbf{r}_1$ and $\mathbf{r}_2$ are the constants of the
reciprocal lattice. The vectors $\mathbf{r}_1$ and $\mathbf{r}_2$,
like $\mathbf{n}_1$ and $\mathbf{n}_2$, are no longer parallel to
the orthogonal $\omega_x$, $\omega_y$-axis, and this introduces a
mixture of $m$ and $k$ sums in the $\omega_y$ functional
dependence. For the convolution theorem, F$(\pmb\omega)$ is
given by:
\begin{equation}
\mathrm{F}(\pmb\omega)=\mathrm{G}(\pmb\omega)\,\Phi(\pmb\omega),
\label{eq:convolut}
\end{equation}
where G$(\pmb\omega)$ is the FT of g$(\mathbf{x})$ and
$\Phi(\pmb\omega)$ is the FT of $\varphi(\mathbf{x})$.
Thus, S$_{\pmb\varepsilon}(\pmb\omega)$ can be split
into three parts: One is dependent on the position of the signal
distribution COG , one is dependent on detector form, and one is
dependent on the signal distribution. Many analytical results can
be deduced by this form. Let us examine the consequences of
Eq.~(\ref{eq:poisson}) on the normalization conservation:
\begin{equation}\label{eq:normaliz}
\begin{aligned}
\mathrm{S}_{\pmb\varepsilon}(0)=&\frac{1}{\tau_{1}\tau_{2}}
\sum_{m,k}\mathrm{e}^{i(m\mathbf{r}_{1}+k\mathbf{r}_{2})\cdot\pmb\varepsilon}
\mathrm{G}(-m\mathbf{r}_{1}-k\mathbf{r}_{2})\,\Phi(-m\mathbf{r}_{1}-k\mathbf{r}_{2})\\
\\
\mathrm{S}_{\pmb\varepsilon}(0)=&1\ \ \ \forall\,
\pmb\varepsilon \ \ \ (\Phi(0)=1).
\end{aligned}
\end{equation}
Due to the linear independence of the exponential function in
$\pmb\varepsilon$  and the arbitrariness of
$\Phi(\pmb\omega)$, the only solution of
Eq.~(\ref{eq:normaliz}) is:
\begin{equation}\label{eq:Gzero}
\mathrm{G}(-m\mathbf{r}_{1}- k\mathbf{r}_{2})= \tau_{1}
\tau_{2}\,\delta_{m,0}\,\delta_{k,0}\ .
\end{equation}

This property introduces a drastic simplification in the
derivation of $\mathbf{r}_g$;  in fact, nonzero results are
obtained by deriving the exponential and G$(\pmb\omega)$.
The partial derivatives of $\Phi(\pmb\omega)$ disappear due
to Eq.~(\ref{eq:Gzero}) which suppresses those calculated outside
the origin $(\pmb\omega=0)$. In the origin, the first
partial derivatives of $\Phi(\pmb\omega)$ are zero for
Eq.~(\ref{eq:uno}). Hence, $\mathbf{r}_g$ becomes:
\begin{equation}\label{eq:COGx}
\mathbf{r}_{g}=\pmb\varepsilon+\frac{i}{\tau_{1}\tau_{2}}\sum_{m,k}\mathrm{e}^{i(m\mathbf{r}_{1}+k\mathbf{r}_{2})
\cdot\pmb\varepsilon}\,
\mathrm{\mathbf{G}}_{\pmb\omega}(-m\mathbf{r}_{1}-k\mathbf{r}_{2})\,
\Phi(-m\mathbf{r}_{1}-k\mathbf{r}_{2}),
\end{equation}
where
$\mathrm{\mathbf{G}}_{\pmb\omega}(\mathbf{a})=($G$_x(\mathbf{a})\,$;$\,$G$_y(\mathbf{a}))$
is the vector of the partial derivatives of
G$(\pmb\omega+\mathbf{a})$ with respect to $\omega_x$ and
$\omega_y$ and taking the limits $\pmb\omega\rightarrow 0$.
Equation~(\ref{eq:COGx}) expresses the functional dependence of
the COG on its parameters and the difference with respect to the
true position. Even if it looks complex, its use is quite simple.
Once the forms of the elementary detector---happily few---have
been defined and calculated for all their FTs and partial
derivatives, one can easily introduce various types of signal
distributions limited only by the complexity of their expressions.
In ~\cite{landi01}, we showed a method to calculate the FT of very
complex signal distributions such as those generated by an
em-shower propagating in a homogeneous medium.

From Eq.~(\ref{eq:COGx}), we can calculate the average square
error of the COG; the $\pmb\varepsilon$-integration of the
exponential function on the detector g$(\mathbf{x})$ has the same
result as an integration of $(x_{g}-\varepsilon_{1})^2$ on a
rectangle of size $\tau_1$ and $\tau_2$. Thanks to
Eq.~(\ref{eq:Gzero}), the $\pmb\varepsilon$-integration can
be transformed in the FT of g$(\mathbf{x})$ calculated at
$\pmb\omega$-values that are differences of the exponents
of Eq.~(\ref{eq:COGx}). In this case, Eq.~(\ref{eq:Gzero})
implements the orthogonality among exponential functions on a
finite domain g$(\mathbf{x})$. Writing the integral of
$(x_g-\varepsilon_1)^2$ over a detector with the function
g$(\mathbf{x})$, we obtain the Parseval identity:
\begin{equation}\label{eq:squareerr}
\int_{\mathbb{R}^{2}}\frac{\mathrm{d}\pmb\varepsilon}{\tau_{1}\tau_{2}}
\mathrm{g}(\pmb\varepsilon)(x_{g}-\varepsilon_{1})^2=
\frac{1}{(\tau_{1}\tau_{2})^{2}}\sum_{m,k}
\big|\mathrm{G}_{x}(-m\mathbf{r}_{1}-k\mathbf{r}_{2})\Big |^{2}\,
\Big |\Phi(-m\mathbf{r}_{1}-k\mathbf{r}_{2})\Big |^{2},
\end{equation}
and identically for $(y_{g}-\varepsilon_{2})^2$ where the only
difference is a change of G$_{x}$ with G$_{y}$. These results,
which can be obtained from our general definitions, will be
applied in selected cases of parallelograms, hexagons, and
triangles.
\section{Parallelograms, Hexagons, Triangles}
\subsection{Parallelograms}

Let us consider parallelogram detectors of the dimensions and with
the bending in Fig.~\ref{fig:due}. We assume that no crosstalk
exists among neighboring detectors, that there are no dead spaces
inbetween, and that their efficiency is constant everywhere. Then,
G$^{p}(\omega_{x},\omega_{y})$ is given by the integral of the
spatial shape g$^{p}(x,y)$ that can be expressed with the interval
function $\Pi(x)$ ($\Pi(x)=1$ for $|x|<1/2$ and $\Pi(x)=0$ for
$|x|\geq 1/2$). This gives
g$^{p}(x,y)=\Pi(y/\tau_{2})\Pi[(x-y/\alpha)/\tau_{1}]$, and its FT
is expressed by:
\begin{align*}
&\mathrm{G}^{p}(\omega_{x},\omega_{y})=\tau_{1}\tau_{2}\,\mathrm{sinc}(\frac{\omega_{x}\tau_{1}}{2})
\,\mathrm{sinc}\big((\omega_{y}+\frac{\omega_{x}}{\alpha})\frac{\tau_{2}}{2}\big )\\
\\
&\mathrm{sinc}(x)=\frac{\sin(x)}{x}\ .
\end{align*}
G$^{p}(\omega_{x},\omega_{y})$ saves the normalization, the system
has no energy loss, and Eq.~(\ref{eq:Gzero}) is easily verified:
\begin{equation*}
\mathrm{G}^{p}(-m\mathbf{r}_{1}-k\mathbf{r}_{2})=\mathrm{G}^{p}(-\frac{2m\pi}{\tau_{1}};\frac{2m\pi}{\tau_{1}\alpha}-
\frac{2k\pi}{\tau_{2}})= \tau_{1}
\tau_{2}\,\delta_{m,0}\,\delta_{k,0}\ .
\end{equation*}
The calculation of G$^{p}_{x,y}(-m\mathbf{r}_{1}-k\mathbf{r}_{2})$
is similarly straightforward and almost always gives zero except
in the following cases:
\begin{align*}
&\mathrm{G}^{p}_{x}(-m\mathbf{r}_{1}-k\mathbf{r}_{2})=
-\frac{\tau_{1}\tau_{2}}{2\pi}\big[\frac{\tau_{1}(-1)^{m}}{m}\delta_{k,0}+\frac{\tau_{2}(-1)^{k}}{\alpha
k}\delta_{m,0}\big]\ \ \ m\neq k,\\
&\mathrm{G}^{p}_{y}(-m\mathbf{r}_{1}-k\mathbf{r}_{2})=
-\frac{\tau_{1}{\tau_{2}}^{2}}{2\pi}\frac{(-1)^{k}}{k}\delta_{m,0}\
\ \ k\neq 0.
\end{align*}
$x_{g}$ and $y_{g}$ are given by:
\begin{align} \label{eq:COGxpar}
x_{g}=\varepsilon_{1}&+\frac{\tau_{1}}{\pi}
\sum_{m=1}^{+\infty}\frac{(-1)^{m}}{m}\mathrm{Imag}\big[
\mathrm{e}^{i\frac{2m\pi}{\tau_{1}}(\varepsilon_{1}-\frac{\varepsilon_{2}}{\alpha})}
\Phi(-m\mathbf{r}_{1})\big]\\
&+\frac{(y_{g}-\varepsilon_{2})}{\alpha}
\nonumber\\
\nonumber\\
y_{g}=\varepsilon_{2}&+\frac{\tau_{2}}{\pi}\sum_{k=1}^{+\infty}\frac{(-1)^{k}}{k}\mathrm{Imag}\big[
\mathrm{e}^{i\frac{2k\pi}{\tau_{2}}\varepsilon_{2}}\Phi(-k\mathbf{r}_{2})\big]
\label{eq:COGypar}
\end{align}
where only the reality of $\varphi(\mathbf{x})$ is used. As
symmetric $\varphi(\mathbf{x})$ has a real and symmetric
$\Phi(\pmb\omega)$, the imaginary part in
Eqs.~(\ref{eq:COGxpar},\ref{eq:COGypar}) is limited to the
exponentials, and gives the $sinus$ function of the respective
argument.

The effect of the parallelogram bending is evident in
Eqs.~(\ref{eq:COGxpar},\ref{eq:COGypar}); the $x_g$
systematic-error contains the
\begin{figure}[h!]
\begin{center}
\includegraphics[scale=0.6]{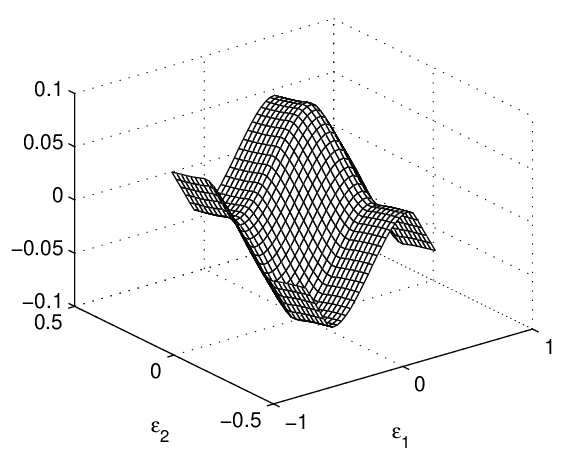}
\includegraphics[scale=0.6]{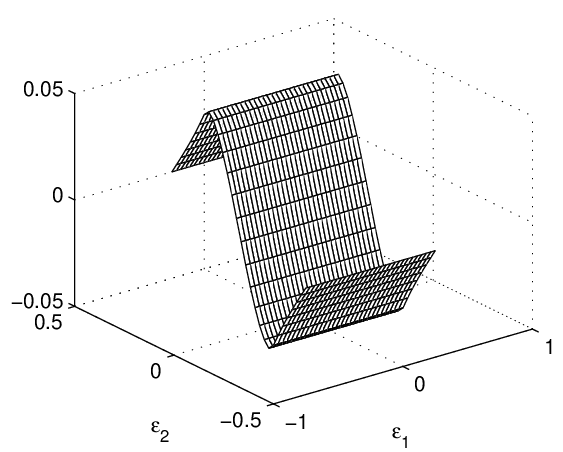}
\end{center}
\caption{\em Map of the discretization error
$x_{g}-\varepsilon_{1}$ (left) and $y_{g}-\varepsilon_{2}$ (right)
for an array of parallelogram detectors with $\tau_{1}=\tau_{2}=1$
and $\alpha=2$. The signal distribution is a disk with radius
$R=1.5\tau$. } \label{fig:tre}
\end{figure}
\begin{figure}[h!]
\begin{center}
\includegraphics[scale=0.8]{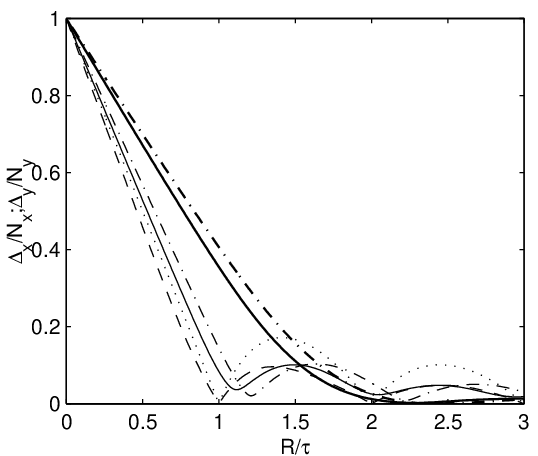}
\end{center}
\caption{\em Average square errors of the array in
Fig.~\ref{fig:tre} for a set of signal distributions with radius
$R$, normalized to the square errors of a Dirac
$\delta$-distribution. The dashed line is the $\Delta_{x_{g}}$ for
rectangles, the dotted lines is $\Delta_{y_{g}}$. The solid line
is $\Delta_{x_{g}}$ for disks, the dashed-dotted line is
$\Delta_{y_{g}}$. The thick lines are the errors for conic-like
form given by the convolution of two disks.}
 \label{fig:quattro}
\end{figure}
$y_g/\alpha$ and $\varepsilon_2/\alpha$ dependence, thus, when
$\alpha\rightarrow\infty$, the two COG errors decouple and $x_g$
has no dependence on $\varepsilon_2$. Hence, it becomes identical
to that calculated in~\cite{landi01}. In all the other cases,
mixed dependence on $\varepsilon_1$ and $\varepsilon_2$ is to be
expected. Applying Eq.~(\ref{eq:squareerr}) with
g$^{p}(\pmb\varepsilon)$ in place of
g$(\pmb\varepsilon)$, we find the average square errors:
\begin{align}\label{eq:Delta2x}
\Delta_{x}^{2}=&\big(\frac{\tau_{1}^{2}}{2\pi^{2}}\big)\
\sum_{m=1}^{+\infty}\frac{1}{m^{2}}\big|\Phi(-m\mathbf{r}_{1})\big|^{2}+
\frac{\Delta_{y}^{2}}{\alpha^{2}}\\
\nonumber\\
\Delta_{y}^{2}=&\big(\frac{\tau_{2}^{2}}{2\pi^{2}}\big)\
\sum_{k=1}^{+\infty}\frac{1}{k^{2}}\big|\Phi(-k\mathbf{r}_{2})\big|^{2}.\label{eq:Delta2y}
\end{align}
Figure~\ref{fig:tre} illustrates the two-dimensional map of the
discretization errors of $x_g$ and $y_g$ of
Eqs.~(\ref{eq:COGxpar},\ref{eq:COGypar}) for a disk-shape signal
distribution of radius $R=1.5\, \tau$. Here we can appreciate the
differences of $x_g$ and $y_g$: $(x_g-\varepsilon_1)$ shows a
simultaneous dependence on $\varepsilon_1$ and $\varepsilon_2$,
and $(y_g-\varepsilon_2)$ does not depend on $\varepsilon_1$. This
is typical of all the parallelogram detectors.

Figure~\ref{fig:quattro} plots the average errors in
Eqs.~(\ref{eq:Delta2x},\ref{eq:Delta2y}) with $R$ increasing from
zero to $3\tau$ for various forms of signal distributions. The
errors are normalized to the value of the error for a Dirac
$\delta$-function signal distribution
($N_{x}^{2}=\tau_{1}^{2}/12+N_{y}^{2}/\alpha$ and
$N_{y}^{2}=\tau_{2}^{2}/12$).

The thin dashed and dotted lines refer to a square signal
distribution: The zeros are given by the $sinus$ functions present
in the FT. The thin solid and dashed-dotted lines are the errors
(on $x_{g}$ and $y_{g}$) of a disk. The origin of the
diffraction-like aspect of the average errors is given by the
zeros of the J$_{1}(\omega)$ (Bessel function of the first kind)
which eliminates the $m=1$ or $k=1$ terms of (\ref{eq:Delta2x})
and (\ref{eq:Delta2y}), leaving only smaller higher order terms.

The thick solid and dashed-dotted lines are the errors of a
conic-like signal distribution given by the convolution of two
identical disks. Its FT is the square of the FT of a disk. We can
see that, above $R/\tau=2$, the COG discretization error
practically disappears. This is due to the first double zeros of
J$_{1}(\omega)^{2}$ whose effects tends to overlap, thereby almost
completely suppressing the COG discretization error for this
signal distribution. This approximate cancellation is interesting
for its efficiency.

\subsection{Hexagons}

As stated above, the sampling along the lines of a set of
parallelograms is able to handle other types of detectors as well.
\begin{figure}[h!]
\begin{center}
\includegraphics[scale=0.8]{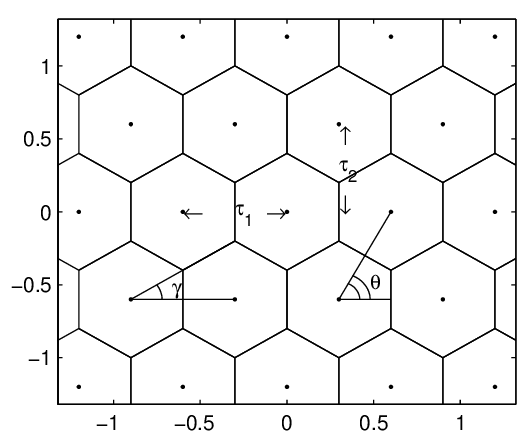}
\end{center}
\caption{\em Array of hexagonal detectors, $\tau_1$ and $\tau_2$
are the x-distance and y-distance of the centers of two
neighboring detectors. The angle $\theta$ is the bending of the
parallelogram as in Fig.~\ref{fig:due}, here
$\tan(\theta)=2\tau_{2}/\tau_{1}$. The angle $\gamma$ is the
inclination of the upper and lower sides of the hexagons. }
 \label{fig:cinque}
\end{figure}
Figure~\ref{fig:cinque} illustrates a hexagonal array showing the
parallelogram sampling strategy. An example of a detector with a
mosaic of hexagons is the 61-pixels HPD of DEP~\cite{DEP}. We
shall limit ourselves to hexagons with two sides parallel to the
$y$-axis. Other types such as those with two sides inclined with
respect to the $y$-axis would be allowed, but, since we do not
know of any detector of this type, we shall ignore them.

As is evident in Fig.~\ref{fig:cinque},
$\alpha=\tan(\theta)=\frac{2\tau_{2}}{\tau_{1}}$, and only
$\tau_{1}$ and $\tau_{2}$ are the free parameters defining the
array; any other parameter can be reduced to them. The angle
$\gamma$ is formed by the inclined lata with respect to the
$x$-axis. It turns out that
$\beta=\tan(\gamma)=\frac{2\tau_{2}}{3\tau_{1}}$, and now
$\mathbf{n}_1=(\tau_1\,;\,0)$ and
$\mathbf{n}_2=(\tau_1/2\,;\,\tau_2)$. Regular hexagons are
obtained with $\gamma=\frac{\pi}{6}$ so
$\frac{\tau_{2}}{\tau_{1}}=\frac{\sqrt{3}}{2}$. To apply the
equations in Section 2, we need the FT of a hexagon, which can be
calculated by:
\begin{equation*}
\mathrm{G}^{h}(\omega_{x};\omega_{y})=\int_{-\frac{\tau_{1}}{2}}^{+\frac{\tau_{1}}{2}}\mathrm{d}
x \int_{\mathrm{y}_{1}(x)}^{\mathrm{y}_{2}(x) }\mathrm{d}
y\,\mathrm{e}^{-i(\omega_{x}x+\omega_{y}y)}
\end{equation*}
where y$_{1}(x)$ and y$_{2}(x)$ are the two set of lines:
\begin{align*}
\mathrm{y}_{1}(x)&=-\beta x-\frac{2}{3}\tau_{2}  \ \ \ \
&\mathrm{y}_{1}(x)=\beta x-\frac{2}{3}\tau_{2}\\
&\ \ \ \ \ x<0 \ \ \ \ &\ \ \ \ \ x\geq 0\\
\mathrm{y}_{2}(x)&=\ \beta x+\frac{2}{3}\tau_{2}  \ \ \ \quad
&\mathrm{y}_{2}(x)=-\beta x+\frac{2}{3}\tau_{2}.
\end{align*}
With some effort, G$^{h}(\omega_{x};\omega_{y})$ assumes the form
(one among the many possible):
\begin{align*}
\mathrm{G}^{h}(\omega_{x};\omega_{y})=\frac{\tau_{1}}{\omega_{y}}\big[&
\sin(\frac{\omega_{x}\tau_{1}}{4}+\frac{\omega_{y}\tau_{2}}{2})
\mathrm{sinc}(\frac{\omega_{x}\tau_{1}}{4}-\frac{\omega_{y}\tau_{2}}{6})-\\
&\sin(\frac{\omega_{x}\tau_{1}}{4}-\frac{\omega_{y}\tau_{2}}{2})
\mathrm{sinc}(\frac{\omega_{x}\tau_{1}}{4}+\frac{\omega_{y}\tau_{2}}{6})\big]\
.
\end{align*}
This form allows easy verification of Eq.~(\ref{eq:Gzero}):
\begin{equation*}
{\underset{\substack{\pmb\omega \rightarrow
0}}{\mathrm{lim}}}\mathrm{G}^{h}(\pmb\omega-m\mathbf{r}_{1}-k\mathbf{r}_{2})=
\tau_{1} \tau_{2} \delta_{m,0}\ \delta_{k,0}\ .
\end{equation*}
Now, $\mathbf{r}_1=(2\pi/\tau_1\,;\,-\pi/\tau_2)$ and
$\mathbf{r}_2=(0\,;\,2\pi/\tau_2)$ are reintroduced, and, as
expected, this hexagon array saves the normalization of
$\varphi(\mathbf{x})$.

The calculation of G$_{x}^{h}$ and G$_{y}^{h}$ is a bit more
complicated, and some help is required of {\textsc {
mathematica}}~\cite{mathematica}. An easy form which coincides
with the partial derivative at $m,k\in \mathbb{Z}^2$ (as is our
concern) can be derived:
\begin{align*}
{\underset{\substack{\pmb\omega\rightarrow
0}}{\mathrm{lim}}}\mathrm{G}_{x}^{h}
(\pmb\omega-m\mathbf{r}_{1}-k\mathbf{r}_{2})&=\frac{9m\tau_{1}^{2}\tau_{2}\sin\big(2(m+k)\frac{\pi}{3}\big)}
{4(m-2k)(2m-k)(m+k)\pi^{2}}\ \ \ m\neq 2k\ \ 2m\neq k\ \ m\neq -k\\
&=-\frac{\tau_{1}^{2}\tau_{2}}{6k\pi}\ \ \ \quad m=2k\\
&=-\frac{\tau_{1}^{2}\tau_{2}}{12m\pi}\ \ \ \ k=2m\ \ \ k=-m\ .
\end{align*}
Using a similar procedure, we can obtain a simplified form of the
partial derivative of G$^{h}(\pmb\omega)$ with respect to
$\omega_y$ for $m,k\in \mathbb{Z}^2$:
\begin{align*}
{\underset{\substack{\pmb\omega\rightarrow
0}}{\mathrm{lim}}}\mathrm{G}_{y}^{h}
(\pmb\omega-m\mathbf{r}_{1}-k\mathbf{r}_{2})&=\frac{-3\tau_{1}\tau_{2}^{2}\sin\big(2(m+k)\frac{\pi}{3}\big)}
{2(2m-k)(m+k)\pi^{2}}\ \ \ 2m\neq k\ \ m\neq -k\\
&=-\frac{\tau_{1}\tau_{2}^{2}}{6m\pi}\ \ \ \quad k=2m\\
&=\ \ \frac{\tau_{1}\tau_{2}^{2}}{6m\pi}\ \ \ \quad k=-m
\end{align*}
For $k=0$ and $m=0$, G$_{x}^{h}(0)$ and G$_{y}^{h}(0)$ are the
hexagon COG-coordinates, and these are zero for our choice of the
reference system origin.

Selecting a signal distribution, we can use Eq.~(\ref{eq:COGx}) to
calculate the COG discretization error for the hexagonal detector
array. The center $(\pmb\varepsilon)$ of the signal
distribution must cover a hexagon, and, due to the periodicity of
the array, the error identically reproduces for all the other
hexagons.
\begin{figure}[h!]
\begin{center}
\includegraphics[scale=0.6]{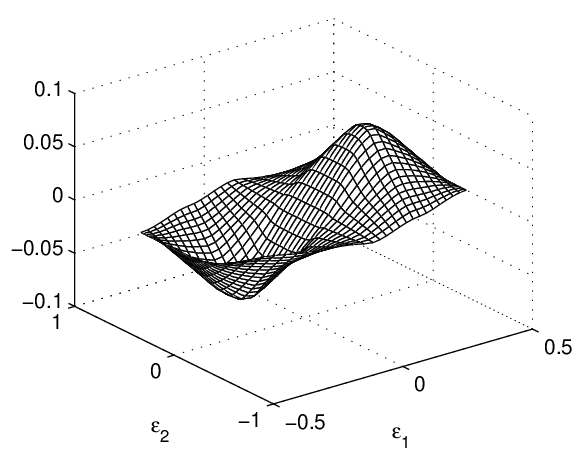}
\includegraphics[scale=0.6]{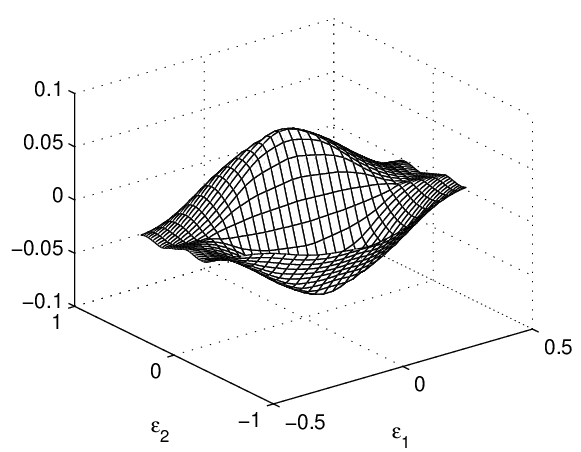}
\end{center}
\caption{\em Map of the discretization error
$x_{g}-\varepsilon_{1}$ (left) and $y_{g}-\varepsilon_{2}$ (right)
for an array of hexagonal detectors with $\tau_{1}=\tau_{2}=1$.
The signal distribution is a disk with radius $R=1.5\tau$. }
 \label{fig:sei}
\end{figure}
The two errors are shown in Fig.~\ref{fig:sei}. They are evidently
different with respect to the parallelogram array in
Fig.~\ref{fig:tre}. Now, the $(x_g-\varepsilon_1)$ and
$(y_g-\varepsilon_2)$ simultaneously depend on $\varepsilon_1$ and
$\varepsilon_2$. The effect of the hexagon's symmetric points
eliminates the discretization error on the vertices and in the
middle. Being less symmetric, the parallelograms have a larger
discretization error.
\begin{figure}[h!]
\begin{center}
\includegraphics[scale=0.8]{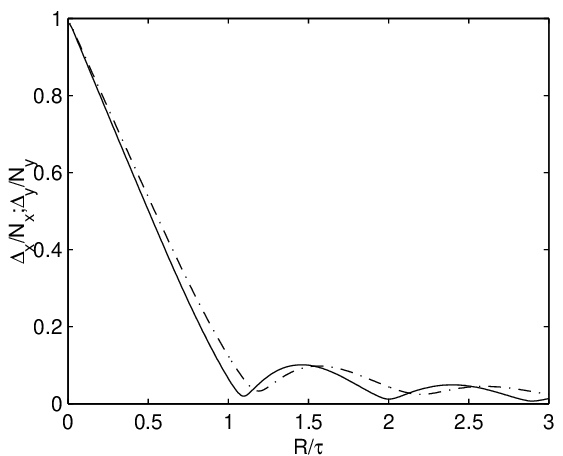}
\end{center}
\caption{\em Average square errors of an array of hexagonal
detectors ($\tau_{1}=\tau_{2}=1$) for a disk signal distributions
with radius $R$ normalized to the square error of a Dirac
$\delta$-distribution on a hexagon. } \label{fig:otto}
\end{figure}

The average square errors can be calculated with
Eq.~(\ref{eq:squareerr}). Figure~\ref{fig:otto} illustrates the
results of its application which in this case is normalized to the
errors of a Dirac $\delta$-function on a hexagon
(N$_{x}^{2}=5\tau_{1}^{2}/72$ and N$_{y}^{2}=5\tau_{2}^{2}/54$).
The plots in Fig.~\ref{fig:otto} greatly resemble those in
Fig.~\ref{fig:quattro}. This is due to the normalization which
always initializes the plots to one. The normalization constants
are smaller for the hexagon compared to the parallelogram. As in
Fig.~\ref{fig:quattro}, the minima are given by the zeros of the
J$_{1}(\omega)$, the FT of a disk. The noticeable effect of the
zeros assures that the higher-order terms of the double infinite
series of Eq.~(\ref{eq:squareerr}) contribute only modestly to the
results. As the other types of signal distributions in
Fig.~\ref{fig:quattro} resemble them, they are not illustrated.

\subsection{Triangles}

Detectors with triangular sections  are rarely used in high energy
physics, an exception is the Crystal Ball~\cite{crystball}. The
Crystal Ball calorimeter is composed of $672$ $NaI$ crystals in
the form of a truncated pyramid with a triangular base. Each
crystal points to the center of the sphere which circumscribes the
calorimeter. The crystal dimensions and the two-dimensional
arrangement are chosen so that the crystals fit inside two sealed
hemispherical containers.
\begin{figure}[h!]
\begin{center}
\includegraphics[scale=0.8]{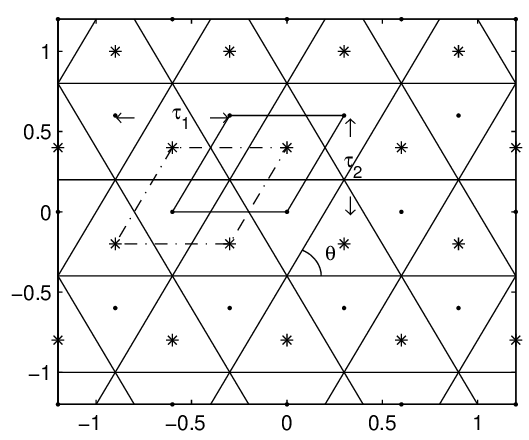}
\end{center}
\caption{\em Array of triangular detectors: $\tau_1$ is the
$x$-distance of the COG of two neighboring detectors, $\tau_2$ is
the $y$-distance, $\theta$ is the bending angle of the inclined
sides. Triangles with a point in the COG are sampled together with
the sampling cell indicated by the solid-line parallelogram.
Triangles with asterisk in the COG are sampled together and the
dashed-dotted parallelogram is their sampling cell.}
\label{fig:nove}
\end{figure}
The spherical geometry can be approximated by our planar
distribution only on a small region.

The sampling differs greatly from the cases considered so far.
First, we have to observe (Fig.~\ref{fig:nove}) that two different
types of triangles are involved: One with a vertex pointing toward
the $x$-axis (indicated in Fig.~\ref{fig:nove} by a dot in the
COG) and one oriented in the opposite direction (indicated in
Fig.~\ref{fig:nove} by an asterisk in the COG). Therefore, two
samplings are required. The periodicity of the samplings is
identical, and a parallelogram is the elementary cell. The
sampling cell is shifted a fixed amount with respect to the other,
and two different convolutions
f$_{\pmb\varepsilon}(\mathbf{x})$ are sampled. Let us call
f$_{1}(\mathbf{x})$ the first convolution with the triangle
pointing downward and f$_{2}(\mathbf{x})$ the convolution with the
triangle pointing upward:
\begin{equation}\label{eq:gconv}
\begin{aligned}
&\mathrm{f}_{1}(\mathbf{x})=\int_{\mathbb{R}^{2}}\mathrm{d}\mathbf{x}'\,
\mathrm{g}_{1}(\mathbf{x}-\mathbf{x}')\,\varphi(\mathbf{x}')\,
, \\
&\mathrm{f}_{2}(\mathbf{x})=\int_{\mathbb{R}^{2}}\mathrm{d}\mathbf{x}'\,
\mathrm{g}_{2}(\mathbf{x}-\mathbf{x}')\,\varphi(\mathbf{x}')\, .
\end{aligned}
\end{equation}
We will only consider isosceles triangles. From
Fig.~\ref{fig:nove}, we can see that
$\alpha=\tan\theta=2\tau_{2}/\tau_{1}$.

Let us define the samplings. The first is generated by a set of
products of Dirac $\delta$-functions centered on the points
$\{n\tau_{1}+l\tau_{1}/2;l\tau_{2}\}, n,l\in\mathbb{Z}^2 $. The
second is generated by a set of products of Dirac
$\delta$-functions centered on the points shifted by $\Delta_{2}$
in the $y$-direction
$\{n\tau_{1}+l\tau_{2}/\alpha;l\tau_{2}+\Delta_{2}\},
n,l\in\mathbb{Z}^2 $ where $\Delta_{2}=\frac{2}{3}\tau_{2}$ (and
$\pmb\Delta_{2}=(0\,;\,\Delta_2)$) is the shift of the COG
of the second type of triangle from that of the first type
centered in the origin (Fig.~\ref{fig:nove}). With these
positions, s$_{\pmb\varepsilon}(\mathbf{x})$ becomes:
\begin{equation}\label{eq:gsampl}
\begin{split}
\mathrm{s}_{\pmb\varepsilon}(\mathbf{x})=\sum_{n,l}\
&\big[\
\delta(\mathbf{x}-n\mathbf{n}_{1}-l\mathbf{n}_{2})\,\mathrm{f}_{1}(n\mathbf{n}_{1}+l\mathbf{n}_{2}-\pmb\varepsilon)+\\
&\delta(\mathbf{x}-n\mathbf{n}_{1}-l\mathbf{n}_{2}-\pmb\Delta_{2})\mathrm{f}_{2}(n\mathbf{n}_{1}+l\mathbf{n}_{2}
+\pmb\Delta_{2}-\pmb\varepsilon)\big].
\end{split}
\end{equation}
The first term accounts for the first set of samplings; the second
term is the shifted sampling on the second type of triangles.

Up to now, we have paid no attention to the difference in the
definition of the detector forms and the function g$(\mathbf{x})$.
We considered them identical, a simplification allowed by the
symmetry of the forms used (rectangles, parallelograms, and
hexagons). Since the triangles are not symmetric with respect to
their COGs, i.e., g$_{1,2}(x,y)\neq $g$_{1,2}(-x,-y)$, we only
have a symmetry with respect to the $y$-axis, i.e., g$_{1,2}(x,y)=
$g$_{1,2}(-x,y)$). This difference must be highlighted and
isolated to avoid incorrect results. The double sampling implies
phase differences which must be recovered by the phase differences
of the FTs of the triangles which are no longer real. The detector
form can be indicated by the function:
\begin{equation}
z=\mathrm{g}_{d}(x-x_{0},y-y_{0}) \label{eq:gfunc}
\end{equation}
where $z$ is the surface generated by g$_{d}$ for the downward
triangle (g$_u$ for the upward). It has a constant value or it is
zero and is centered in $x_0$ and $y_{0}$ where $x_0$, $y_0$ are
the coordinates of the detector COG. Comparing
Eq.~(\ref{eq:gfunc}) with Eqs.~(\ref{eq:gconv}) and
(\ref{eq:gsampl}), we find the differences: Eq.~(\ref{eq:gconv})
contains g$_{1}(x_{0}-x';y_{0}-y')$ where $x_0$ and $y_0$ are now
the coordinates of the detector COG, yielding:
\begin{equation}\label{g=g}
\mathrm{g}_{1}(x_{0}-x;y_{0}-y)=\mathrm{g}_{d}(-(x-x_{0});-(y-y_{0})).
\end{equation}
For symmetric detectors, this difference is irrelevant, but for
triangles it represents a complete inversion. In fact, having
defined g$_{d}(\mathbf{x})$ as the first type of triangle, we find
g$_{d}(-\mathbf{x})$ as the second type. If we incorrectly take
for g$_{1}(\mathbf{x})$ the function g$_{d}(\mathbf{x})$ in the
convolution (\ref{eq:gconv}), the function g$_{1}(-\mathbf{x})$ is
actually used. This implies an exchange between the two types of
triangles;  Eq.~(\ref{eq:Gzero}) is no longer verified due to the
overlap of the triangles and the holes left in the plane. Let us
calculate the form of G$_{d}(\pmb\omega)$ the FT of
g$_{d}(\mathbf{x})$, and
S$_{\pmb\varepsilon}(\pmb\omega)$ the FT of
s$_{\pmb\varepsilon}(\mathbf{x})$. The function
g$_{d}(x,y)$ can be expressed by the interval function $\Pi(x)$ as
defined in Section~3.1:
\begin{equation}\label{eq:FTtri}
\begin{aligned}
&\mathrm{g}_{d}(x,y)=\Pi(\frac{y+\frac{\tau_{2}}{6}}{\tau_{2}})
\Pi(\frac{x}{\frac{2}{\alpha}(y+\frac{2}{3}\tau_{2})})\\
\\
&\mathrm{G}_{d}(\omega_{x};\omega_{y})=\int_{-\frac{2}{3}\tau_{2}}^{\frac{1}{3}\tau_{2}}\mathrm{d}
y \int_{-(y+\frac{2}{3}\tau_{2})
\frac{1}{\alpha}}^{(y+\frac{2}{3}\tau_{2})\frac{1}{\alpha}}\mathrm{d} x
\mathrm{e}^{-i(\omega_{x}x+\omega_{y}y)}\ .
\end{aligned}
\end{equation}
Among the forms of G$_{d}(\omega_{x};\omega_{y})$ we shall
indicate:
\begin{equation}\label{eq:Gtri}
\begin{split}
\mathrm{G}_{d}(\omega_{x};\omega_{y})= \frac{4\tau_{1}\tau_{2}\
\mathrm{e}^{-i\frac{\omega_{y}\tau_{2}}{3}}}
{(4\omega_{y}^{2}\tau_{2}^{2}-\omega_{x}^{2}\tau_{1}^{2})}\big[
\cos(\frac{\omega_{x}\tau_{1}}{2})-
\mathrm{e}^{i\omega_{y}\tau_{2}}+i\omega_{y}\tau_{2} \
\mathrm{sinc}(\frac{\omega_{x}\tau_{1}}{2})\big]\,.
\end{split}
\end{equation}
Due to the double sampling, the calculation of the FT of
s$_{\pmb\varepsilon}(\mathbf{x})$ becomes more complicated.
With two applications of the procedure described for
Eq.~(\ref{eq:poisson}), the Poisson relation for this case can be
derived as:
\begin{align}
\mathrm{S}_{\pmb\varepsilon}(\pmb\omega)=&
\sum_{m,k}\mathrm{e}^{-i(\pmb\omega-\mathbf{L}_{m\,k})\cdot\pmb\varepsilon}
\big[\mathrm{G}_{d}^{*}(\pmb\omega-\mathbf{L}_{m\,k})
+\mathrm{e}^{i(m-2k)\frac{2}{3}\pi}
\mathrm{G}_{d}(\pmb\omega-\mathbf{L}_{m\,k})
\big]\Phi(\pmb\omega-\mathbf{L}_{m\,k})\nonumber\\
\label{eq:FTstria}\\
 \mathbf{L}_{m\,k}=&
m\mathbf{r}_{1}+k\mathbf{r}_{2}=\big(\frac{2m\pi}{\tau_{1}}
\,;\,\frac{\pi}{\tau_{2}}(2k-m)\big)\nonumber
\end{align}
where G$_{d}^{*}(\pmb\omega)$ is the complex conjugate of
G$_{d}(\pmb\omega)$ defined by Eq.~(\ref{eq:FTtri}) and is
the FT of g$_{2}(x,y)$ of Eq.~(\ref{eq:gconv}) as explained above.

We can verify that G$_{d}(-\mathbf{L}_{m\,k})$ becomes zero for
integers $m\neq 0$ and $k\neq 0$; nonzero contributions are
obtained when $m=0$ and $k\neq 0$ or $m\neq 0$ and $k=0$. These
terms are cancelled out by G$^{*}_{d}$. The phase factor behind
G$_{d}$ in Eq.~(\ref{eq:FTstria}) is crucial for the cancellation:
Only the term with $m=0$ and $k=0$ survives, and
Eq.~(\ref{eq:Gzero}) is verified. The sampling arrangement of
Eq.~(\ref{eq:gsampl}) effectively tessellates the plane with
triangular detectors.

The combination of two triangles, one pointing upward and the
other downward, forms a parallelogram. This property simplifies
the partial derivatives of
S$_{\pmb\varepsilon}(\pmb\omega)$, and the nonzero
elements needed to calculate the COG are:
\begin{equation}\label{eq:Gxtria}
\begin{aligned}
&\mathrm{G}^{t}(\pmb\omega)=\mathrm{G}_{d}^{*}(\pmb\omega)+
\mathrm{e}^{i\pi\frac{2}{3}(m-2
k)}\mathrm{G}_{d}(\pmb\omega)\\
\\
&{\underset{\substack{\pmb\omega\rightarrow 0\\k=0}}{\mathrm{lim}}}
\mathrm{G}_{x}^{t}(\pmb\omega-\mathbf{L}_{m\,0})=\\
&{\underset{\substack{\pmb\omega\rightarrow
0\\k=m}}{\mathrm{lim}}}
\mathrm{G}_{x}^{t}(\pmb\omega-\mathbf{L}_{m\, m})=
\frac{-\tau_{1}^{2}\tau_{2}}{4}\frac{\mathrm{e}^{-i\frac{4}{3}m\pi}}{m\pi}
\end{aligned}
\end{equation}
for $m$ integer and $m\neq 0$. Similarly, G$_{y}^{t}$ is:
\begin{equation}\label{eq:Gytria}
\begin{aligned}
&{\underset{\substack{\pmb\omega\rightarrow
0\\k=0}}{\mathrm{lim}}}
\mathrm{G}_{y}^{t}(\pmb\omega-\mathbf{L}_{m\,0})=\\
&{\underset{\substack{\pmb\omega\rightarrow
0\\k=m}}{\mathrm{lim}}}
\mathrm{G}_{y}^{t}(\pmb\omega-\mathbf{L}_{m\, m})=
\frac{\tau_{1}\tau_{2}^{2}}{6}\frac{\mathrm{e}^{-i\frac{2}{3}m\pi}}{m\pi}
\end{aligned}
\end{equation}
for $m\neq 0$ and $m$ integer. Using Eqs.~(\ref{eq:Gxtria}) and
(\ref{eq:Gytria}), we can calculate $x_g$ and $y_g$, which turn
out to be:
\begin{align}
x_{g}=&\varepsilon_{1}+\frac{\tau_{1}}{2\pi}\sum_{m=1}^{+\infty}\frac{1}{m}\mathrm{Imag}
\Big\{\mathrm{e}^{im\pi(\frac{2\varepsilon_{1}}{\tau_{1}}-\frac{4}{3})}\big[
\mathrm{e}^{-i\frac{m\pi}{\tau_{2}}\varepsilon_{2}}
\Phi(-\mathbf{L}_{m\, 0})+
\mathrm{e}^{i\frac{m\pi}{\tau_{2}}\varepsilon_{2}}
\Phi(-\mathbf{L}_{m\,m})\big]\Big\}\nonumber\\
\label{eq:COGtria}\\
y_{g}=&\varepsilon_{2}-\frac{\tau_{2}}{3\pi}\sum_{m=1}^{+\infty}\frac{1}{m}\mathrm{Imag}
\Big\{\mathrm{e}^{im\pi(\frac{2\varepsilon_{1}}{\tau_{1}}-\frac{2}{3})}\big[
\mathrm{e}^{-i\frac{m\pi}{\tau_{2}}\varepsilon_{2}}
\Phi(-\mathbf{L}_{m\,0})+
\mathrm{e}^{i\frac{m\pi}{\tau_{2}}\varepsilon_{2}}
\Phi(-\mathbf{L}_{m\,m})\big]\Big\}.\nonumber
\end{align}
\begin{figure}[h!]
\begin{center}
\includegraphics[scale=0.6]{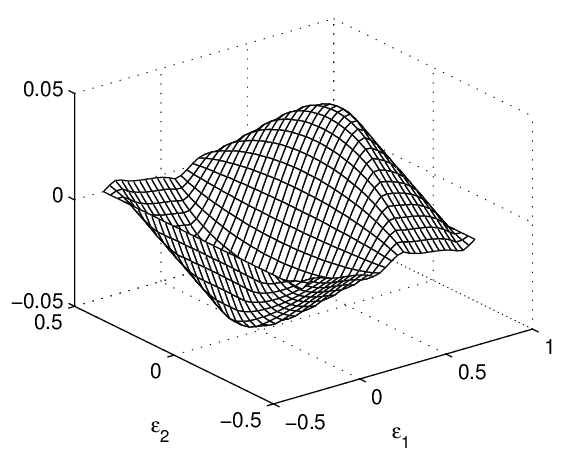}
\includegraphics[scale=0.6]{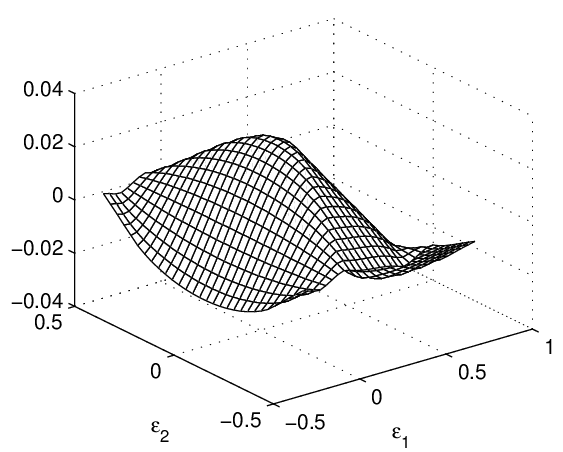}
\end{center}
\caption{\em Map of the discretization error
$x_{g}-\varepsilon_{1}$ (left) and $y_{g}-\varepsilon_{2}$ (right)
for an array of triangular detectors as in Fig.~\ref{fig:nove},
but calculated on a rectangular-elementary cell. The signal
distribution is again a disk with radius $R=1.5\tau$. }
 \label{fig:dieci}
\end{figure}
Although the two sums in Eq. (\ref{eq:COGtria}) look identical,
they actually differ by a phase factor contained in the first
exponential. For Eq. (\ref{eq:squareerr}), this phase difference
is irrelevant for the square mean errors $\Delta_{x}^{2}$ and
$\Delta_{y}^{2}$ that become proportional, and are given by:
\begin{equation}\label{eq:Delta2tria}
\begin{aligned}
&\Delta_{x}^{2}=\tau_{1}^{2}\sum_{m=1}^{+\infty}\frac{1}{8m^{2}\pi^{2}}
\Big[\big|\Phi(-\mathbf{L}_{m\,0})\big|^{2}
+\big|\Phi(-\mathbf{L}_{m\,m})\big|^{2}\Big]\\
&\Delta_{y}^{2}=\frac{4\tau_{2}^{2}}{9\tau_{1}^{2}}\Delta_{x}^{2}\
.
\end{aligned}
\end{equation}
\begin{figure}[h!]
\begin{center}
\includegraphics[scale=0.8]{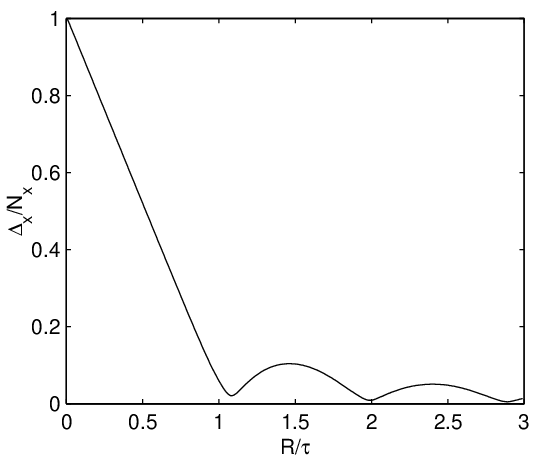}
\end{center}
\caption{\em Average square errors of an array of triangular
detectors ($\tau_{1}=\tau_{2}=1$) for a disk signal distribution
with radius $R$ ranging from zero to $3\tau_1$ normalized to the
square error of a Dirac $\delta$-distribution.} \label{fig:undici}
\end{figure}
In the plots of $(x_{g}-\varepsilon_{1})$ and of
$(y_{g}-\varepsilon_{2})$ shown in Fig.~\ref{fig:dieci}, the
elementary cell is taken as a rectangle because, in the case of
axisymmetric signal distribution, a few lines parallel to the
$y$-axis have zero discretization error. Figure~\ref{fig:undici}
shows $\Delta_{x,y}$. From the second
equation~(\ref{eq:Delta2tria}), we can see that the normalized
errors $\Delta_{x,y}$ are now identical, the errors are always
proportional, and the normalization reabsorbs the constant. Here,
too, the average square error is suppressed for the disk signal
distribution of special sizes, which is always produced by the
zeros of the J$_1(\omega)$ Bessel function (FT of a disk).

\section{Special Signal Distributions}
\subsection{Signal Distributions Free of Discretization
Error}

In the previous sections, we illustrated the properties of some
detector arrays. Our selection only partially covers the richness
of possible configurations allowed by the two-dimensional geometry
as compared with the one-dimensional geometry
. We encounter a similar large set of configurations for signal
distributions free of discretization error.

The easiest case of band-limited functions has some additional
properties. From Eq. (\ref{eq:squareerr}), we can see that
$\Phi(\pmb\omega)$ has two different band-limits:
$\Phi(\omega_{x},\omega_{y})=0$ for $|\omega_{x}|\geq
2\pi/\tau_{1}$ and $|\omega_{y}|\geq 2\pi/\tau_{2}$. If the FT of
$\varphi(\mathbf{x})$ has these limitations, $\Delta_{x}^{2}$ and
$\Delta_{y}^{2}$ are always zero, no matter what type of detectors
are considered, granted that the detector array has a periodicity
controlled by parameters $\tau_1$ and $\tau_2$. Roughly speaking,
we can say that, even in two dimensions, the origin of the
discretization error $\Delta_{x,y}$ can be connected to the
aliasing effects due to the improper sampling of the function
f$_{\pmb\varepsilon}(\mathbf{x})$ with overly large
sampling steps for the band-limits (if any) of the function
$\varphi(\mathbf{x})$. The band boundaries which could eliminate
the discretization error are twice those required by the
Wittaker-Kotel'nikov-Shannon (WKS) sampling
theorem~\cite{WKS,libroFT2} for a complete reconstruction of the
function f$_{\pmb\varepsilon}(\mathbf{x})$ from their
sampled values at intervals $\tau_1$ and $\tau_2$.

As demonstrated in Ref.~\cite{libroFT2,mallat}, a band-limited
function cannot be zero on any finite part of the plane. This
property rules out these functions from the set of functions
usually encountered in experimental position measurements. Here,
the functions $\varphi(\mathbf{x})$ become zero after only a few
$\tau_{1}$ or $\tau_2$, and no band-limitations can be invoked or
postulated. The elimination of the COG discretization error can be
obtained from other sources. We can see in some of the previous
figures that  the discretization error disappears for special
forms and special sizes.

In the one-dimensional case, the interval function and all its
convolutions with normalized functions have sizes without
discretization error. In two dimensions, this set is much larger.
It is evident from Eq.~(\ref{eq:squareerr}) that: {\em{All the
signal distributions with the property of Eq.~(\ref{eq:Gzero})
with the same $\{\alpha,\tau_{1},\tau_{2}\}$ as the detector array
are free of COG-discretization error}}. For example, for
$\alpha=2$ we explicitly demonstrate that a hexagonal distribution
has the property (\ref{eq:Gzero}). Therefore, a hexagonal signal
distribution with sizes $\tau_1$ and $\tau_2$ or any of their
multiples (and all the other related parameters as described in
Fig.~\ref{fig:cinque}) has an exact $x,y$-COG on an array of
isosceles triangles with base $\tau_1$ and height $\tau_2$ and
arranged as in Fig.~\ref{fig:nove}. Highly irregular shapes such
as those illustrated in Ref.~\cite{Escher} are prototypes of
signal distributions free of the COG discretization errors. We can
easily prove from the previous equations that even disconnected
signal distributions have an exact COG. For example, any two
triangles (one pointing upward and one downward) in
Fig.~\ref{fig:nove} have exact COGs on the hexagon array in
Fig.~\ref{fig:cinque}. Each triangle has a COG error, but, paired,
the errors cancel out.

Evidently, all the convolutions of arbitrary functions
with functions having the property (\ref{eq:Gzero}) continue to
maintain it and are free of COG-discretization errors. A
rectangular signal distribution of dimensions $\tau_1$ and
$\tau_2$ (or any of their multiples) is free of discretization
error for any array with the same $\tau_{1} $, $\tau_{2} $ and any
value of $\alpha$. We can easily figure out the geometric origin
of the property. An array of rectangles of dimensions
$\{\tau_{1},\tau_{2}\}$, where each line of detectors is shifted
with respect to the previous line of a fixed amount $\sigma$
tessellate the plane. Therefore, Eq.~(\ref{eq:Gzero}) must hold,
and due to $\alpha=\tau_{2}/\sigma$ any type of array with the
values $\{\tau_{1},\tau_{2},\alpha\}$ is easily generated. Even
the parallelogram and triangle arrays  continue to tessellate the
plane after a shift of one detector line with respect to another
and when the shift is repeated in all the detector lines. In the
following subsection, we will describe a few properties of arrays
of shifted rectangles.
\begin{figure}[h!]
\begin{center}
\includegraphics[scale=0.8]{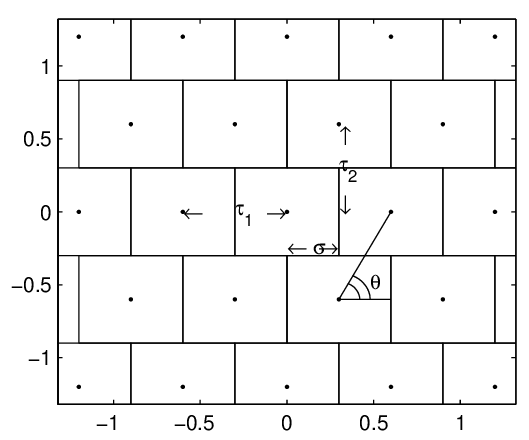}
\end{center}
\caption{\em Array of shifted rectangular detectors. $\tau_1$ and
$\tau_2$ are defined as usual. $\sigma$ is the shift of the
detector centers of a row with respect to the detector centers of
the row below them. Tan$(\theta)=\tau_{2}/\sigma$ is the bending
angle of the line connecting the centers of two rectangles
pertaining to two different rows.} \label{fig:dodici}
\end{figure}

\subsection{Shifted Rectangles}

To exemplify the property of an array of shifted detectors, we
shall describe the easiest one, i.e., that of the array of shifted
rectangles (Fig.~\ref{fig:dodici}). It is evident that, at the
variation of the shift $\sigma$, all the values of
$\alpha=\tan(\theta)=\tau_{2}/\sigma$ are explored, and Eq.
(\ref{eq:sampling})-(\ref{eq:squareerr}) must be applied to the
sampling. The FT of g$^{r}(x,y)$ of a rectangle is given by:
\begin{equation*}
\mathrm{G}^{r}(\omega_{x},\omega_{y})=\tau_{1}\tau_{2}\,\mathrm{sinc}(\frac{\omega_{x}\tau_{1}}{2})
\,\mathrm{sinc}\big(\frac{\omega_{y}\tau_{2}}{2}\big ).
\end{equation*}
Writing Eq.~(\ref{eq:Gzero}) for this case, we get:
\begin{equation}\label{eq:Gr}
\mathrm{G}^{r}(-m\mathbf{r}_{1}-k\mathbf{r}_{2})=\tau_{1}\tau_{2}\,\mathrm{sinc}(-m\pi)
\,\mathrm{sinc}\big(\frac{2m\pi\tau_{2}}{\tau_{1}\alpha}- k\pi\big
)=
 \tau_{1}
\tau_{2}\,\delta_{m,0}\,\delta_{k,0}\ .
\end{equation}
where the first $sinc$-function eliminates any $m$ and $\alpha$
dependence in the second one. The tessellation condition is
verified for any $\alpha$ as expected by naive intuition.
Equation~(\ref{eq:Gr}) is what we need to prove that a rectangular
form of dimensions $\tau_1$ and $\tau_2$ (and all the multiples of
these dimensions) is free of discretization error for any array
with the same $\tau_1$ and $\tau_2$, any $\alpha$ and any form of
detectors which tessellate the plane. We can easily verify that
the same is true for parallelograms and triangles. The hexagons
cannot be shifted.

\subsection{Scaled Signal Distributions}

In discussing Eq.~(\ref{eq:Gzero}), we considered an array of
detectors covering the plane, but among the periodic forms that
satisfy Eq.~(\ref{eq:Gzero}), we have to consider their scaled
forms. For the scaling property of the FT, if we define
$\mathrm{g}_l(\mathbf{x})=\mathrm{g}(\mathbf{x}/l)/l^2$ than
$\mathrm{G}_l(\pmb\omega)=\mathrm{G}(l\pmb\omega)$.
Thus, if $\mathrm{g}(\mathbf{x})$ satisfies Eq.~(\ref{eq:Gzero}),
then $\mathrm{g}_l(\mathbf{x})$ satisfies the same equation for
any positive integer $l$. The normalized signal distribution
$\varphi(\mathbf{x})=\mathrm{g}_l(\mathbf{x})/(\tau_1\tau_2)$ is
free of discretization error. As detectors, the functions
$\mathrm{g}_l(\mathbf{x})$ have crosstalk with the neighboring
detectors.

\subsection{Approximate Results}

Some previous figures of the average square errors report deep
minima for some R-values of disk-shaped signal distribution. These
minima do not reach zero values of the square error, but are
always suppressions of the COG discretization error worthy of
exploration. For this reason, all the plots in
Eq.~(\ref{eq:squareerr}) exhibit the error for the disk signal
distribution. Cell bending $\alpha$ has some effect on the minima
positions. For parallelograms and hexagons, we can see a slight
shift in the minimum for $\Delta_x$ and $\Delta_y$ and $\alpha=2$,
for larger $\alpha$-values the shift tends to disappear as can be
verified from Eqs.~(\ref{eq:Delta2x}) and (\ref{eq:Delta2y}) for
parallelograms. The FT of a disk is $4 J_1(R\omega)/(R\omega)$,
and the discretized values of $\omega_x$ and $\omega_y$ modulate
the effect of the zeros of $J_1(\omega)$ in
Eq.~(\ref{eq:squareerr}), giving a shift of the minimum. The
interesting aspect of this approximate result is its applicability
to almost all the detector arrays as far as $\tau_1\sim \tau_2$.

\section{Crosstalk and Loss}
\subsection{Loss}

If the detectors in the array have generic (nonuniform) crosstalk
or they have some losses, Eq.~(\ref{eq:normaliz}) need some
modification. We must consider:
\begin{equation}
\mathrm{S}_{\pmb\varepsilon}(0)=\sum_{nl}
\mathrm{f}(n\mathbf{n}_{1}+l\mathbf{n}_{2}-\pmb\varepsilon)
\leq 1\ .
\end{equation}
The detector array does not generate a constant (over the plane)
efficiency function, and in some regions the efficiency drops
below one. Now S$_{\pmb\varepsilon}(0)$ retains its
dependence on the position and form of the signal distribution,
and Eq.~(\ref{eq:Gzero}) is no longer valid. To calculate the COG,
we have to use Eq.~(\ref{eq:FTcog}) with the partial derivatives
of $\Phi(\pmb\omega)$:
\begin{equation}\label{eq:39}
\begin{aligned}
&\mathbf{r}_{g}=\pmb\varepsilon+
\frac{i}{\tau_{1}\tau_{2}\mathrm{S}_{\pmb\varepsilon}(0)}\,
\sum_{m,\,k}\mathrm{e}^{i\mathbf{L}_{m\,k}\cdot\,\pmb\varepsilon}[\,
\mathrm{\mathbf{G}}_{\pmb\omega}(-\mathbf{L}_{m\,k})\Phi(-\mathbf{L}_{m\,k})+
\mathrm{G}(-\mathbf{L}_{m\,k})\pmb\Phi_{\pmb\omega}(-\mathbf{L}_{m\,k})\big]\\
\\
&\mathbf{L}_{m\,k}=m\mathbf{r}_{1}+k\mathbf{r}_{2}=
\big(\frac{2m\pi}{\tau_{1}}\,;\,-\frac{2m\pi}{\tau_{1}\alpha}+\frac{2k\pi}{\tau_{2}}\big)\,
.
\end{aligned}
\end{equation}
Here, $\pmb\Phi_{\pmb\omega}=(\Phi_x\,;\,\Phi_y)$ is
the vector of the partial derivatives with respect to $\omega_{x}$
and $\omega_y$ of $\Phi(\omega_{x},\omega_{y})$.
Equation~(\ref{eq:39}) allows some simplification if the signal
distribution is one of the special
\begin{figure}[h!]
\begin{center}
\includegraphics[height=5.6cm,width=7cm]{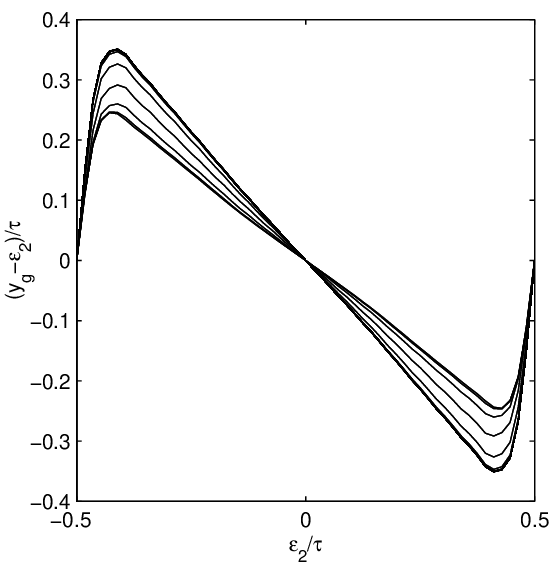}
\end{center}
\caption{\em Set of error curves ($\varepsilon_{2}-\tau_{2}$) for
a signal distribution formed by two disks, one of small radius
(0.15 $\tau$) and considerable height (4.5), and a disk with a
larger radius (1.5 $\tau$) and relative height 1; the detectors
are squares with a region of loss at any border for a range of
$0.025\tau$. } \label{fig:dodiciadd}
\end{figure}
functions discussed in Section 4. If $\varphi(\mathbf{x})$ is a
band-limited function with $\Phi(\omega_{x},\omega_{y})=0$ for
$|\omega_{x}|\geq 2\pi/\tau_{1}$ and $|\omega_{y}\geq
2\pi/\tau_{2}$, the COG discretization error disappears for any
type of loss or crosstalk; in this case,
S$_{\pmb\varepsilon}(0)$ is always constant. If
$\varphi(\mathbf{x})$ is one of the finite support functions which
have no COG error in absence of loss (or crosstalk),
S$_{\pmb\varepsilon}(0)$ is constant for any
g$(\mathbf{x})$, and only the terms depending on the partial
derivatives of $\Phi(\pmb\omega)$ remain in Eq.
(\ref{eq:39}). For functions whose first partial derivatives
pertain to previous class, even the last term of Eq. (\ref{eq:39})
disappears for any type of loss or crosstalk.

Figure~\ref{fig:dodiciadd} shows one of the loss effects, i.e.,
the introduction of a simultaneous dependence on $\varepsilon_1$
and $\varepsilon_2$ in the COG error, even for an array of
detectors that in Section 2 was proved to be dependent on
$\varepsilon_1$ or on $\varepsilon_2$, but not on both.

\subsection{Uniform Crosstalk}

Equations~(\ref{eq:uniform},\ref{eq:normS}) give a definition of
uniform crosstalk as a property of the function g$(\mathbf{x})$:
\begin{equation}\label{eq:uniform1}
\sum_{nl}\mathrm{g}(n\tau_{1}+\frac{l\tau_{2}}{\alpha}-x;l\tau_{2}-y)=1\
\ \ \ \mathit{a.\  e.}
\end{equation}
Let us examine the possibility of verifying or building functions
g$(\mathbf{x})$ with uniform crosstalk. The crosstalk is an effect
of a detector on its neighbors in the array, so the true physical
form of the detectors partially loses its meaning. Theoretically,
this effect can be of any type, even completely independent of the
detector form. In practice, this is probably rare, so we will
retain some of the detector properties. To generate uniform
crosstalk which retains memory of the detector form, the rule is
simple: {\em {Any convolution of detector functions
g$(\mathbf{x})$ of the type considered in Section 4 with an
arbitrary normalized function g$_{2}(\mathbf{x})$, generates
uniform crosstalk}}. This condition allows us to verify Eqs.
(\ref{eq:uniform1}) and (\ref{eq:normS}). If g$_{1}(\mathbf{x})$
is one of the functions g$(\mathbf{x})$ studied in Section 4 and
its corresponding array tessellates the plane for a set of values
$\{\tau_{1},\tau_{2},\alpha\}$, then g$_{1}(\mathbf{x})$
definitely satisfies Eq. (\ref{eq:Gzero}) and likewise any of its
convolutions. It is evident that g$(\mathbf{x})$, convolution of
g$_{1}(\mathbf{x})$ and g$_{2}(\mathbf{x})$ (or more), has no
constant values over the detector surface. In fact, we assumed
that g$_{1}(\mathbf{x})$ tessellate the plane with constant values
over a detector, and  this cannot be true for all its
convolutions. Thus, g$(\mathbf{x})$ can spread outside the
physical boundaries of the detector and thereby generate an
external crosstalk. Contrary to the intuition, this crosstalk does
not induce a signal loss, and can improve the position
reconstruction without deteriorating the energy measurement (if we
are dealing with a calorimeter). Identical properties are valid
for any integer scaling of the dimensions of g$(\mathbf{x})$ and
for any of its convolutions.
\subsection{A Form of the WKS-theorem}

Equation~(\ref{eq:uniform1}) can even be the starting point for
other mathematical extrapolations along the lines described in
Refs.~\cite{WKS,mallat}. We will limit to develop a connection to the
nonseparable form of the WKS-theorem in two dimensions.

A straightforward extension of the WKS-theorem to a
two-dimensional problem is executed with double application to the
$x$ and $y$ variables. In this way, the sampling points are
arranged on an orthogonal lattice, and the orthogonal set of
functions is the tensorial product of the two one-dimensional set.
For construction, each function of the set has a separable
dependence from $x$ and $y$ that can be non optimal for some
problems. Here we have a set of equations that allow a different
expression of the theorem.

To directly use the equations in the previous sections, we will
work in the dual space: The functions in the $\mathbf{x}$-space
are range-limited (the dual form of band-limited) and the
functions in the $\pmb\omega$-space will be reconstructed
by sampling at the appropriate frequencies.

Owing to the definitions we gave, the  functions g$(\mathbf{x})$
for the parallelograms, hexagons, and triangles tessellate the
plane and possess the property:
\begin{equation*}
\mathrm{g}^{2}(\mathbf{x})=\mathrm{g}(\mathbf{x}).
\end{equation*}
For the convolution theorem, this implies:
\begin{equation*}
\int_{\mathbb{R}^{2}}\mathrm{G}(\pmb\omega-\pmb\omega')\mathrm{G}(\pmb\omega'-\pmb\omega'')
\frac{\mathrm{d}
\pmb\omega'}{4\pi^{2}}=\mathrm{G}(\pmb\omega-\pmb\omega'')\,.
\end{equation*}
The reality of g$(\mathbf{x})$ gives
G$(\pmb\omega)=$G$^{*}(-\pmb\omega)$. Substituting
this in the integral and taking
$\pmb\omega=m\mathbf{r}_{1}+l\mathbf{r}_{2}$ and
$\pmb\omega''=n\mathbf{r}_{1}+k\mathbf{r}_{2}$ with $
m,n,l,k\in\mathbb{Z}^4$, Eq~(\ref{eq:Gzero}) imposes the
orthogonality of the set of functions
G$_{m,l}(\pmb\omega)=\mathrm{G}(\pmb\omega-m\mathbf{r}_{1}-l\mathbf{r}_{2})$.
This is a direct consequence of Eq.~(\ref{eq:uniform1}) and the
plane tessellation implemented by g$(\mathbf{x})$ (and obviously
by g$(\mathbf{x})^2$).

For any function $\varphi( \mathbf{x})$ such that $\varphi(
\mathbf{x})\mathrm{g}(\mathbf{x})=\varphi( \mathbf{x})$ (i.e.,for
functions different from zero within the boundaries of
g$(\mathbf{x})$), the FT can be written:
\begin{equation*}
\begin{aligned}
&\Phi(\pmb\omega)=\sum_{m,l\in\,\mathbb{Z}^2}\Phi(m\mathbf{r}_1+l\mathbf{r}_2)
\frac{\mathrm{G}_{m,l}(\pmb\omega)}{\tau_1\tau_2}\\
&\Phi(m\mathbf{r}_1+l\mathbf{r}_2)=\int_{\mathbb{R}^2}
\mathrm{G}^*(\pmb\omega'-m\mathbf{r}_1-l\mathbf{r}_2)\Phi(\pmb\omega')
\frac{\mathrm{d}\pmb\omega'}{4\pi^2}\,.
\end{aligned}
\end{equation*}
Again, given Eq.~(\ref{eq:Gzero}) for G$(\pmb\omega)$,
$\Phi(m\mathbf{r}_1+l\mathbf{r}_2)$ are the samples of the
function $\Phi(\pmb\omega)$ at frequencies
$m\mathbf{r}_1+l\mathbf{r}_2$. The completeness of the set of
functions $\{$G$_{m,l}(\pmb\omega)\}$ is more complicated
to prove, but expressing the functions
G$_{m,l}(\pmb\omega)$ as FT of g$(\mathbf{x})$ and using
two times the formal relation in Eq.~(\ref{eq:poisson}), we can
prove that:
\begin{equation*}
\sum_{m,l\in\mathbb{Z}^2}\mathrm{G}_{m,l}^*(\pmb\omega')\mathrm{G}_{m,l}(\pmb\omega)=
\tau_{1}\tau_{2}\mathrm{G}(\pmb\omega-\pmb\omega')\,.
\end{equation*}
Substituting in the previous equations, the completeness is
verified. The basis G$^p$ is very similar to a double application
of the one-dimensional WKS-theorem, and becomes separable for
$\alpha\rightarrow\infty$. The bases G$^h$ and G$^t$ are
nonseparable in a product of two functions dependent on $\omega_x$
and $\omega_y$, and involve functions whose FTs have support
contained in a hexagon or in a pair of triangles. This approach
provides a physical model for the WKS-theorem as a property of an
array of constant efficiency detectors.

\subsection{Crosstalk Free of the COG Discretization Error: the Ideal Detector}

Uniformity is only one of the crosstalk properties. Evidently,
the signal spreading almost always reduces the COG error, but
some forms of crosstalk are {\em{free of the COG discretization
error for any signal distribution}}. We saw above that it is
possible to select special signal distributions free of the
COG-discretization errors; however, very few choices are possible
in the case of the signal distribution. The detailed physical
process generating the signal is usually beyond the control of
detectors. Conversely, the crosstalk can be under the control of
the detector fabrication and, at least theoretically, some
possibilities exist to converge toward some form of crosstalk as
described. Even if no concrete example can be given, let us
discuss which form of crosstalk is free of the COG discretization
errors for any signal distribution in a two-dimensional geometry.

Previously, we discussed how to generate crosstalk g$(\mathbf{x})$
functions with constant efficiency over the array. With such
g$(\mathbf{x})$, the signal collected by a detector affects its
neighbors. From the point of view of the calculation of
$\mathbf{r}_g$, Eq.~(\ref{eq:COGx}) remains identical. The
uniformity of g$(\mathbf{x})$ as expressed by
Eq.~(\ref{eq:uniform1}), eliminates the partial derivatives of the
function $\Phi(\pmb\omega)$ from the COG expression. A
further condition can be imposed on g$(\mathbf{x})$ to eliminate
even the partial derivatives of G$(\pmb\omega)$. We
supposed above that g$(\mathbf{x})$ was the convolution of
functions g$_{1}(\mathbf{x})$ and g$_{2}(\mathbf{x})$. The first,
g$_1(\mathbf{x})$, was constrained to tessellate the plane with
the same $\{\tau_1,\tau_2,\alpha\}$ of this array, but no
condition was posed on the g$_{2}(\mathbf{x})$. If
g$_{2}(\mathbf{x})$ is constrained to tessellate the plane with
the same $\{\tau_1,\tau_2,\alpha\}$, the partial derivatives
G$(\pmb\omega)$ are zero for $\omega_{x}\rightarrow
-2m\pi/\tau_{1}$ and $\omega_{y}\rightarrow 2\pi(m/\tau_{1}\alpha
-k/\tau_{2})$: i.e.,
\begin{equation}\label{eq:noerr}
\begin{aligned}
&\mathrm{G}_{x}(a,b)=\mathrm{G}_{1x}(a,b)\mathrm{G}_{2}(a,b)+\mathrm{G}_{1}(a,b)\mathrm{G}_{2x}(a,b)\\
&a=-\frac{2m\pi}{\tau_{1}}\ \ \ \ \ \
b=2\pi(\frac{m}{\tau_{1}\alpha}-\frac{k}{\tau_{2}})\\
&\mathrm{G}_{1}(a,b)=\mathrm{G}_{2}(a,b)=0\ \ \ \ \ \
\mathrm{G}_{1x}(0,0)=\mathrm{G}_{2x}(0,0)=0,
\end{aligned}
\end{equation}
and likewise for the $y$-partial derivative. It is evident that,
if g$(\mathbf{x})$ is the convolution of three or more functions
in addition to g$_{1}$ and g$_{2}$, the absence of the COG
discretization error is maintained. In the two-dimensional case,
the set of the simplest forms free of COG errors for any signal
distribution is much larger then in one-dimension. We proved
in~\cite{landi01} that a triangle with base $2\tau$ for g$(x)$ was
the simplest shape for eliminating the COG error. Now, for
$\alpha=2$, we can select two functions among hexagons, triangles,
parallelograms, or shifted rectangles with the same
$\{\tau_{1},\tau_{2}\}$ to cite only the
G$(\omega_{x},\omega_{y})$ we explicitly calculated, but many more
can be invented.
\begin{figure}[h!]
\begin{center}
\includegraphics[scale=0.7]{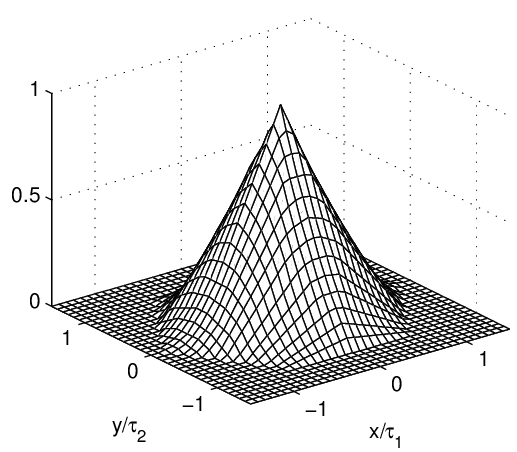}
\includegraphics[scale=0.7]{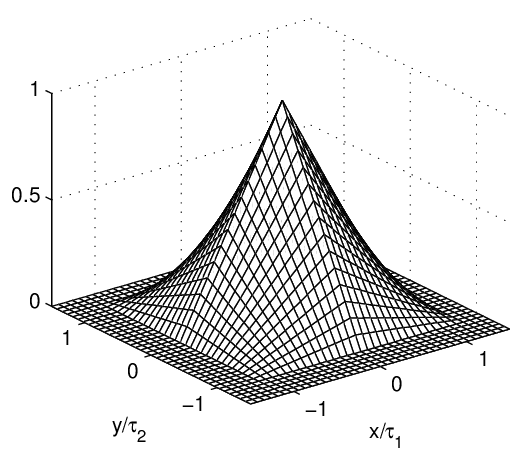}
\end{center}
\caption{\em Convolution of two hexagons of an array with
$\tau_1=\tau_2$ (left). Convolution of two rectangles (right). The
first type of crosstalk eliminates the COG discretization error
for any array of detectors with $\tau_1=\tau_2$ and $\alpha=2$;
the second eliminates it even for any $\alpha$. }
\label{fig:tredici}
\end{figure}

Figure~\ref{fig:tredici} shows two forms of crosstalk. We can see
that the convolution of two constant height hexagonal surface
generates a surface with negligible edges much resembling a cone.
The base is always hexagonal. In the convolution of two constant
height square surfaces, two edges are present. If the two
convolved forms are both sized $\tau_{1}$x$\,\tau_{2}$, their
convolution is $2\tau_{1}$x$\,2\tau_{2}$. These are the smallest
sizes. The convolution of any crosstalk function free of
discretization error with any normalized function saves the
property, but the size of the resulting function is the sum of the
sizes of the two convolved functions.

With a set of steps giving Eq. (\ref{eq:nocog}), it can be proved
that any function g$(x,y)$ with the following property is free of
the COG discretization error for $x_g$ and for $y_g$:
\begin{equation}\label{eq:quadue}
\begin{aligned}
&\sum_{n,l=-\infty}^{+\infty}(n\tau_{1}+\frac{l\tau_{2}}{\alpha}-x)
\ \mathrm{g}(n\tau_{1}+\frac{l\tau_{2}}{\alpha}-x;l\tau_{2}-y)=0\
\ \ &(\text{a.\ e.})\\
&\sum_{n,l=-\infty}^{+\infty}(l\tau_{2}-y) \
\mathrm{g}(n\tau_{1}+\frac{l\tau_{2}}{\alpha}-x;l\tau_{2}-y)=0\ \
\ \ \ \quad\quad &(\text{a.\ e.})\ .
\end{aligned}
\end{equation}

We are unable to give an example of detectors which implement or
attempt to approach this feature. In the one-dimensional case, the
silicon strip detector with floating strips probably approaches a
crosstalk which tends to a form free of discretization error for
any signal distribution.

It is useful to discuss the criticality of condition
(\ref{eq:quadue}) in the presence of a size (or period) mismatch.
For this,
\begin{figure}[h!]
\begin{center}
\includegraphics[scale=0.8]{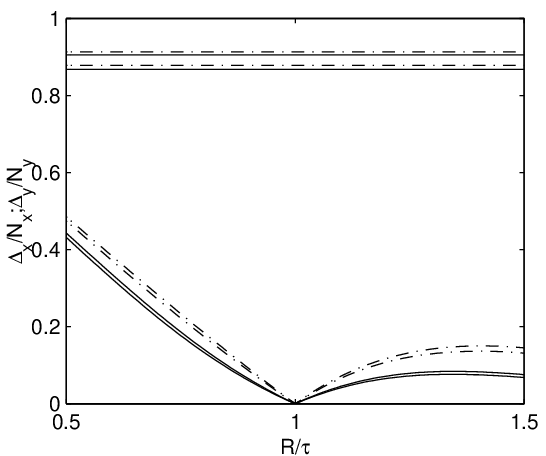}
\end{center}
\caption{\em Effects of the crosstalk on a pair of near
$\delta$-like signals. The upper curves are $\Delta_{x_{g}}$ and
$\Delta_{y_{g}}$ for a square array without crosstalk. The lower
curves are given by the presence of a square crosstalk and ranges
running from $R/\tau=0.5$ up to $R/\tau=1.5$. For $R/\tau=1$, the
error is completely suppressed. } \label{fig:quattordici}
\end{figure}
we construct a simulation with a narrow signal distribution and a
large discretization error on an array with $\tau_{1}=\tau_2=1$,
and a crosstalk function dimensioned around the optimal value. A
dramatic improvement --- even for large discrepancies with respect
to the exact value --- can be seen in Fig.~\ref{fig:quattordici}.
The crosstalk is a square of varying dimensions. This form is
probably of interest for the pixel detector of CMS~\cite{CMS1},
but it is impractical for homogeneous detectors where axisymmetric
shapes are more probable. The smallest size axisymmetric crosstalk
which heavily suppresses the COG discretization error is the disk
of $R/\tau\approx 1.2$ due to the first zero of J$_{1}(\omega)$ FT
of a disk (in this case g$(x,y)$ has a range of $2.2\tau$). Even
if this is not an exact suppression, it looks very efficient and
almost independent of $\alpha$, at least for large (or small)
$\alpha$ values.

Another approximate suppression of the COG error is given by the
conoid form of the crosstalk, which is obtained by the convolution
of two disks. Here the overlap of the zeros of J$_{1}(\omega)^{2}$
almost completely suppresses the COG error beyond $R/\tau\approx
2$. Axial symmetry is granted even in this case, and the
insensibility to the $\alpha$ values is more pronounced than in
the case of a disk. On the basis of these considerations,
Fig.~\ref{fig:quattro} can be read as the effects of various types
of crosstalk on a $\delta$-like signal distribution.

\section{Finite Array of Detectors}
\subsection{Imaging on a Pixel Detector}

Up to now, we have explored an infinite periodic array of
detectors, and the sampling-function periodicity was the key
property in extracting explicit analytical results. With a finite
set of detectors, periodicity disappears and calculation of the
convolutions is unavoidable. But, as proved in
Ref.~\cite{landi01}, the solution to the problem lies in the
finite range of the signal distribution $\varphi(\mathbf{x})$ and
the limited number of detectors ($2$x$3,\ 3$x$3,\ 5$x$5,\dots$).
Without losing the locality property of $\mathbf{r}_g$, we can use
a periodic repetition of $\varphi(\mathbf{x})$ (indicated with
$\varphi^{p}(\mathbf{x})$) or the periodic repetition of the set
of detectors. Since the second looks more involved analytically,
we will use the first approach. The periodic
$\varphi^{p}(\mathbf{x})$ explores the set of detectors, and if
the periods are sufficiently ample, the (periodic) convolution
with the sensor function g$(\mathbf{x})$ coincides with the
convolution of $\varphi(\mathbf{x})$ on a period. As
in~\cite{landi01}, the periodic function $\varphi^{p}(\mathbf{x})$
with periods $T_1$ and $T_2$ is defined as (for  $\,r, s \in
\mathbb{Z}^2$ and with the vectors $ \mathbf{T}_1=(T_1\,;\,0)$,
and $ \mathbf{T}_2=(0\,;\,T_2)$):
\begin{equation*}
\varphi^{p}(\mathbf{x})=\sum_{r,s}
\varphi(\mathbf{x}-r\mathbf{T}_{1}-s\mathbf{T}_{2})\, .
\end{equation*}
The FT of $\varphi^{p}(\mathbf{x})$ (indicated with
$\Phi^{p}(\pmb\omega)$) has components at frequencies that
are multiples of the fundamental frequencies $2\pi/T_{1}$ and
$2\pi/T_{2}$ and introducing the vectors
$\mathbf{R}_1=(2\pi/T_1\,;\,0)$ and $
\mathbf{R}_2=(0\,;\,2\pi/T_2)$:
\begin{equation}\label{eq:periodFI}
\Phi^{p}(\pmb\omega)=\frac{(2\pi)^2}{T_{1}T_{2}}\sum_{m,k}
\delta(\pmb\omega-m\mathbf{R}_{1}-k\mathbf{R}_{2})\,
\Phi(m\,R_{1};k\,R_{2}),
\end{equation}
where the amplitudes $\Phi(m\,R_{1};k\,R_{2})$ are the values of
the FT of $\varphi(\mathbf{x})$ taken at the corresponding values.
We will assume that $T_1$ and $T_2$ are larger than the region
involved in the sampling (i.e., the set of sensors are well
contained in a rectangle $T_1$x$T_2$), so we can write:
\begin{equation*}
\begin{aligned}
&\mathrm{f}^{p}(n\mathbf{n}_{1}+l\mathbf{n}_{2}-\pmb\varepsilon)=
\mathrm{f}(n\mathbf{n}_{1}+l\mathbf{n}_{2}-\pmb\varepsilon)\
\ \ \text{n,l}\ \epsilon\ I_{0}\ \ \
\varepsilon_{1},\varepsilon_{2}\ \epsilon\ C_{1}\,. \\
&\mathrm{f}^{p}(\mathbf{x})=\sum_{m,k}\mathrm{e}^{i(m\,\mathbf{R}_{1}+k\,\mathbf{R}_{2})\cdot\,\mathbf{x}}
\mathrm{G}(mR_{1}\,;\,kR_{2})\Phi(mR_{1}\,;kR_{2})\frac{1}{T_{1}T_{2}}\,.
\end{aligned}
\end{equation*}
Now, the set $I_0$ of the detector indexes $n$ and $l$ is some
subset of integers and $\pmb\varepsilon$ covers a detector.
We define a function s$^{p}_{\pmb\varepsilon}(\mathbf{x})$
as:
\begin{equation*}
\mathrm{s}^{p}_{\pmb\varepsilon}(\mathbf{x})=\sum_{n,l\,\in\,I_{0}}\delta(\mathbf{x}-n\mathbf{n}_{1}-l\mathbf{n}_{2})
\,\mathrm{f}^{p}(\mathbf{x}-\pmb\varepsilon)
\end{equation*}
where s$^{p}_{\pmb\varepsilon}(\mathbf{x})$ is expressed as
sum of products of two $\mathbf{x}$-functions. The FT of
s$^{p}_{\pmb\varepsilon}(\mathbf{x})$ is the convolution of
the FTs of the two functions. However, f$^{p}(\mathbf{x})$ is
periodic and has a FT similar to that of Eq. (\ref{eq:periodFI}).
This eliminates the integrations in the convolution and blocks the
integration variables to the values of their $\delta$-functions.
Then, S$^{p}_{\pmb\varepsilon}(\pmb\omega)$ becomes:
\begin{equation}\label{eq:Szero}
\begin{aligned}
\mathrm{S}^{p}_{\pmb\varepsilon}(\pmb\omega)=&\int_{\mathbb{R}^{2}}
\frac{d\pmb\omega}{4\pi^{2}}
\mathrm{H}(\pmb\omega-\pmb\omega')\mathrm{F}^{p}(\pmb\omega')\\
\mathrm{S}^{p}_{\pmb\varepsilon}(\pmb\omega)=&\frac{1}{T_{1}T_{2}}\sum_{m,k\,\in\,\mathbb{Z}^{2}}\
\sum_{n,l\,\in\,I_{0}}\mathrm{e}^{-i(n\mathbf{n}_{1}+l\mathbf{n}_{2})
(\pmb\omega-m\mathbf{R}_{1}-k\mathbf{R}_{2})}\\
&\mathrm{e}^{-i(m\mathbf{R}_{1}+k\mathbf{R}_{2})\cdot\,\pmb\varepsilon}
\mathrm{G}(mR_{1};kR_{2})\Phi(mR_{1};kR_{2})\ ,
\end{aligned}
\end{equation}
where H$(\pmb\omega)$ is the FT of the set of Dirac
$\delta$-functions that define the detector positions.

Applying Eqs. (\ref{eq:COGx}) to Eq. (\ref{eq:Szero}), we can
calculate the two components of $\mathbf{r}_g$. These equations
can be easily generalized to a finite array of different detectors
as the type discussed in~\cite{klanner}, or to a subset of
detectors of Penrose tessellation~\cite{Penrose}. Each type of
detector is introduced with its proper (different)
f$^{p}_{n,l}(\mathbf{x})$ produced by a g$_{n,l}(\mathbf{x})$
function convolving the same signal distribution. The
g$_{n,l}(\mathbf{x})$ functions account for the properties of the
detectors located in the points labelled by the indexes $\{n,l\}$.
Then, we must modify Eq.~(\ref{eq:Szero}) accordingly, introducing
functions G$_{n,l}(\pmb\omega)$, FT of
g$_{n,l}(\mathbf{x})$, that contain the shape, efficiency etc. of
the detectors with their COGs in position $\{x_n,y_l\}$.

\subsection{3x3 and 5x5 Detector Arrays}

Let us describe the $3$x$3$ and $5$x$5$ arrays of rectangular
detectors of the same type. In Eq. (\ref{eq:Szero}), the two sets
of indexes $n$ and $l$ run on the integers $\{-1,\ 0,\ 1\}$, for
the $3$x$3$ array, and $\{-2, -1,\ 0,\ 1,\ 2\}$ for the $5$x$5$.
H$^{5\mathrm{x}5}(\pmb\omega)$ and
H$^{3\mathrm{x}3}(\pmb\omega)$ becomes:
\begin{equation}\label{H55}
\begin{aligned}
\mathrm{H}^{5\mathrm{x}5}(\pmb\omega)=&[2\cos(2\omega_x\tau_1)+2\cos(2\omega_x\tau_1)+1]
[2\cos(2\omega_y\tau_2)+2\cos(2\omega_y\tau_2)+1]\\
\mathrm{H}^{3\mathrm{x}3}(\pmb\omega)=&[2\cos(2\omega_x\tau_1)+1]
[2\cos(2\omega_y\tau_2)+1]
\end{aligned}
\end{equation}
and $x_g$ for a $3$x$3$ array becomes:
\begin{equation}\label{Xg3x3}
\begin{aligned}
x^{3\mathrm{x}3}_{g}=\frac{i}{T_1
T_2}\sum_{m,k}&2\tau_1\sin(mR_1\tau_1)
\big[2\cos(kR_2\tau_2)+1\big]\,\mathrm{e}^{-i(m\mathbf{R}_1+k\mathbf{R}_2)\,
\cdot\,\pmb\varepsilon}\,\\
&\mathrm{G}(mR_1;kR_2)\, \Phi(mR_1;kR_2)\,.
\end{aligned}
\end{equation}
Similar expressions can be easily obtained for $y_g$ and for a
$5$x$5$ array. As discussed in~\cite{landi01}, a set of
discontinuities are present in
$\mathbf{r}_g(\pmb\varepsilon)$ if the signal distribution
extends beyond the limits of the array. In both cases, the
discontinuities are located at the borders of the central
detector. Near the borders, the suppressed tails of the signal
distributions tend to differ, thereby inducing an error in the
COGs. Beyond the borders, a different set of detectors are
inserted in the algorithm, and the error changes sign, generating
a
\begin{figure}[h!]
\begin{center}
\includegraphics[scale=0.6]{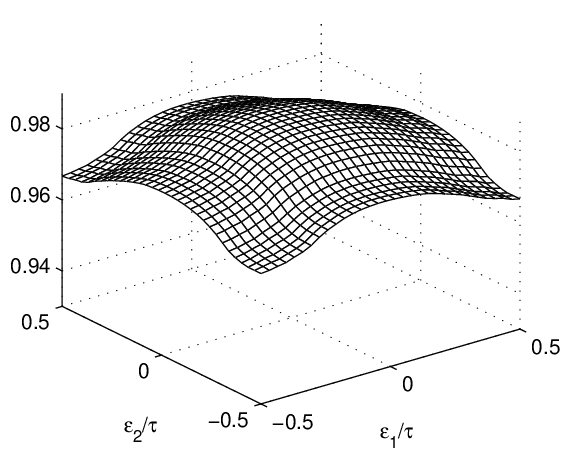}
\includegraphics[scale=0.6]{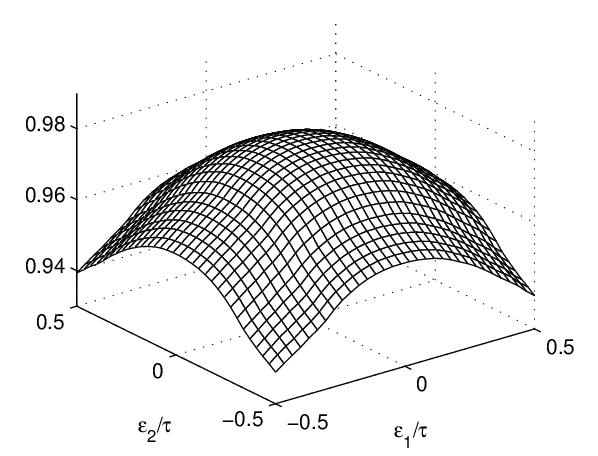}
\end{center}
\caption{\em The surface on the left is the efficiency for energy
collection of the $5$x$5$ algorithm (i.e.,
S$^{5\mathrm{x}5}(0,0)$). The surface on the right is the
efficiency of the $3$x$3$ algorithm (S$^{3\mathrm{x}3}(0,0)$). }
\label{fig:sedici}
\end{figure}
discontinuity.

To show how the $3$x$3$ and $5$x$5$ algorithms work, we will apply
them to a detector similar to the CMS em calorimeter~\cite{CMS2}
with a signal distribution generated as described
in~\cite{landi01} for a set of tracks. In~\cite{landi01}, we
explored the form of the average signal distribution generated by
a set of random tracks whose parameters where tuned to reproduce
the average lateral signal distribution of an em shower as
extracted from the data in Ref.~\cite{lednev}. The shower's axis
is parallel to the detector's $z$-axis, so the tracks' useful
parameters do not contain their $z$-origin. In the CMS em
calorimeter, the crystals have a nonprojective geometry with a
$3^{\circ}$x$\,3^{\circ}$ off-center pointing. In the simulation
the shower axis is bent respect to the detector's $z$-axis, and
the longitudinal distribution of the track origins must be
introduced. This can be produced by adding a $z$-dependent
probability distribution of the track origins whose average
reproduces the average longitudinal energy distribution of a $200$
GeV photon-shower as parameterized in~\cite{particledatabook}. The
set of tracks are globally bent $3^{\circ}$ in $y$- and in
$x$-direction to simulate the off-axis distortion of the
CMS-crystals. The average over $150$ distributions is used to
calculate  the efficiency surfaces of the $5$x$5$ and $3$x$3$
energy reconstruction algorithms. It should be stressed that this
is not a Monte Carlo simulation, but a toy model that allows easy
calculation of S$^{5\mathrm{x}5}(0,0)$ and
S$^{3\mathrm{x}3}(0,0)$. Most likely, the average signal
distribution is realistic enough for our needs: In fact, the
signal distribution is filtered by a low-pass filter and very few
details are relevant to this simulation.

The bending of the em-shower axis renders the signal distribution
slightly asymmetric. The COG shower does not coincide with the
impact point of the photon on the face of the central crystal. To
reconstruct the impact point, further assumptions are required.
Our natural reference point is the COG signal distribution, which
sets the maximum of the efficiency plots in the center of the
detector.  The range of allowed $\pmb\varepsilon$-values is
limited by the central-detector boundaries.

Figure~\ref{fig:sedici} shows the efficiency plots of algorithms
$5$x$5$ and $3$x$3$. There is evidently greater signal loss and
greater suppression of the signal tails in the $3$x$3$ array. The
marked reductions at the borders of the $5$x$5$ array are due to
the loss at the boundaries between crystals. This loss is
simulated as a dead space whose length is the fraction of
effective radiation length of the interspersed material. In these
plots, we neglect an effect of the asymmetry of the signal
distribution that, at a pair of corners, moves the maximum
collected energy on a nearby crystal not containing the COG
shower.
\begin{figure}[h!]
\begin{center}
\includegraphics[scale=0.70]{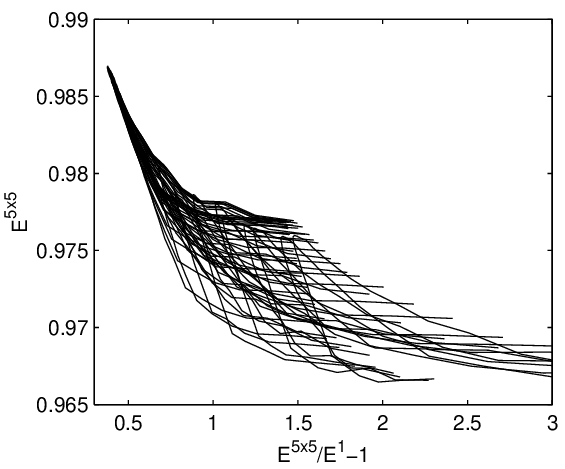}
\caption{\em Energy collected by the 5x5 array E$\,^{5x5}$in
function of the quantity E$\,^{5x5}/$E$^1-1$. This plot cannot be
reduced to a line as is usually done for the energy
reconstruction.} \label{fig:diciassettea}
\end{center}
\end{figure}
\begin{figure}[h!]
\begin{center}
\includegraphics[scale=0.70]{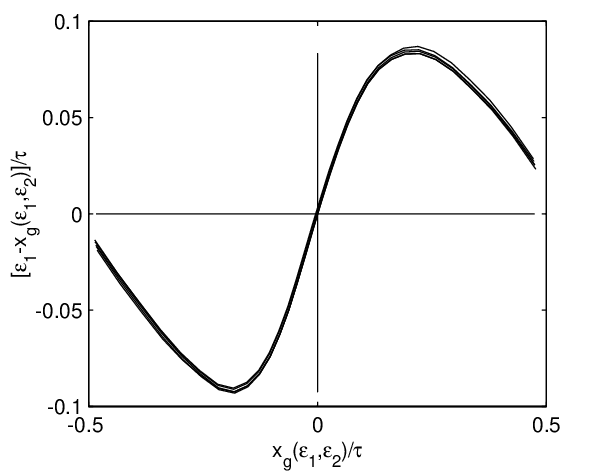}
\caption{\em Set of plots for various values of $\varepsilon_2$ of
COG error $\varepsilon_1-x_g$, each shown in function of $x_g$.}
\label{fig:diciassetteb}
\end{center}
\end{figure}

The efficiency surfaces of Fig.~\ref{fig:sedici} show a decrease
at the borders. The decrease is usually recovered by {\it{ad hoc}}
reconstruction methods. One method~\cite{CMS2} is to consider the
plot of E$\,^{5\mathrm{x}5}$ in relation to
E$\,^{5\mathrm{x}5}$/E$^1-1$ and fit the resulting distribution
(E$^1$ is the energy collected by the central detector). It is
evident from Fig.~\ref{fig:diciassettea} that this function cannot
be fit to a line without introducing an effective broadening of
the energy resolution. The functions E$^{5\mathrm{x}5}$ and E$^1$
are surfaces and have a complex correspondence. It is our guess
that accounting for them could improve the energy reconstruction.

The COG error $(\varepsilon_1-x_g)$ in relation to $x_g$ is
illustrated in Fig.~\ref{fig:diciassetteb} for an array of $3$x$3$
detectors. This quantity is a function of $\varepsilon_1$ and
$\varepsilon_2$, but, contrary to expectations, the dependence on
$\varepsilon_2$ is negligible. This almost decoupling of the $x_g$
error from $\varepsilon_2$ and $y_g$ error from $\varepsilon_1$
allows each error to be corrected independently with the methods
discussed in~\cite{landi01}. A slight dependence of $x_g$ on
$\varepsilon_2$ is probably introduced by the border regions where
the maximum signal detector is not that containing the COG shower.
These regions are at the corners of the central detector, where
the signal spreads almost evenly over four detectors, and the COG
error is very small.

\subsection{Triangular Detectors}

This subsection is devoted to describe the energy and position
reconstructions in the Crystal Ball (CB)
detector~\cite{crystball}, a first-generation em calorimeter still
in operation at the BNL~\cite{CBBNL} which is applied to the study
of barionic resonances. This application illustrates the
effectiveness of Eqs.~(\ref{eq:periodFI},\ref{eq:Szero}) to a
complex geometry such as that of the CB. As mentioned in Section
3.3, the CB calorimeter is composed of 672 optically isolated
{\it{NaI(Tl)}} crystals. Each crystal is shaped as a truncated
triangular pyramid, 40.6 cm long, pointing toward the interaction
point, with 5.1 cm sides at the inner end and 12.7 cm sides at the
outer end. The crystal setup is based on the geometry of an
icosahedron having an equatorial plane that allows the separation
of the ball into two hemispheres for easily opening. Each of the
twenty main triangular faces of the icosahedron ($major\
triangles$) is divided into four $minor\ triangles$, each one
consisting of nine individual crystals. Some crystals are removed
at two sides of the equatorial plane to make room for the beam
pipe.

Some details are ignored in the simulation. Having to work on a
plane, we project a restricted set of triangular detectors. The
significant differences in size of the top and bottom are
neglected; each detector's size will be that measured at the
shower maximum ($40$ cm from the center of the CB for electron
showers and $0.8$ radiation lengths more for photon showers). The
energy reconstruction algorithm (as used long ago in the CB
experiment in DESY at the DORIS II accelerator~\cite{CBmio}) sums
all the energy released in 12 crystals around the crystal with the
maximum energy of the cluster and the maximum itself (E$_{13}$).

The triangles' relative positions in the array are crucial. The
$n$ and $l$ values for the downward triangles to be used in
Eq.~(\ref{eq:Szero}) are: $[(0,1)$ $(-1,1)$ $(-1,0)$ $(0,0)$
$(1,0)$ $(0,-1)$ $(1,-1)]$. The upward triangles have $n$ and $l$
given by: $[(-1,0)$ $(0,0)$ $(1,0)$ $(0,-1)$ $(1,-1)$ $(1,-2)]$.
It should be recalled that the origin of the sampling for the
upward triangles is shifted by $2\,\tau_2/3$ and $\alpha$ is
$2\,\tau_2/\tau_1$. The triangles are equilateral with
$\tau_1=3.37$ radiation lengths and high $\tau_2$. Using
Eq.~(\ref{eq:FTcog}) in Eq.~(\ref{eq:Szero}), we obtain
$\mathbf{r}_g(\pmb\varepsilon)$. Due to the lack of
symmetry, we must carefully account for the relative phases of the
complex quantities involved.
\begin{figure}[h!]
\begin{center}
\includegraphics[scale=0.7]{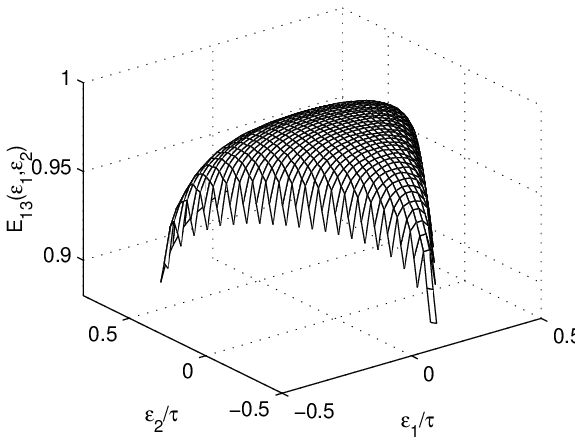}
\includegraphics[scale=0.7]{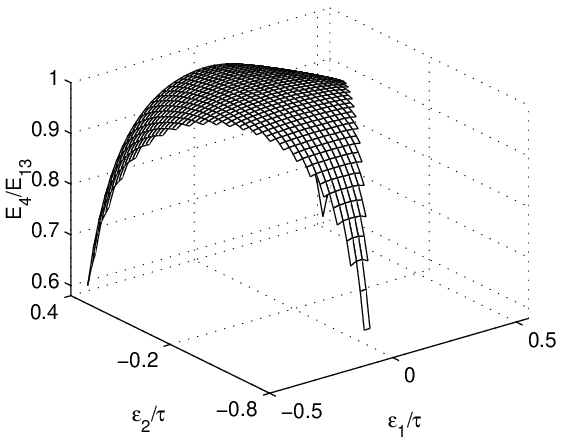}
\end{center}
\caption{\em The surface on the left is
S$_{\varepsilon_1,\varepsilon_2}(0,0)$ for the array of 13
crystals. $\varepsilon_1$ and $\varepsilon_2$ explore the central
crystal. The surface on the right is the ration E$_{4}/$E$_{13}$.
Here, the efficiency drop at the borders of a crystal disappears,
while the drop at the corners increases.}\label{fig:diciotto}
\end{figure}

We apply Eqs.~(\ref{eq:periodFI},\ref{eq:Szero}) to the set of
triangles, remembering the differences of the upward and downward
triangles as discussed in Section 3.3. We must use
Eq.~(\ref{eq:Gtri}) in Eq.~(\ref{eq:Szero}) for the upward
triangles and its complex conjugate for the downward triangles.
The average signal distribution is the model distribution
generated as described in~\cite{landi01} to reproduce the lateral
signal distribution of an em shower. Here no $z$-distribution of
the tracks is needed due to the coincidence of the CB center with
the interaction point (projective geometry). The loss at the
crystal borders is probably underestimated, since the model
assumes that all the photons convert. Theoretically, the
projective geometry allows photons to escape detection with an
efficiency dropping near zero.

The efficiency plot in Fig.~\ref{fig:diciotto} is calculated for
the range of $\pmb\varepsilon$-value covering the central
triangle. As expected, the plot is a surface, and surfaces are the
ratios with other partial sums of the collected energy. For
example, histograms of E$_4/$E$_{13}$ were used in cuts to reduce
the $\pi_0$ contamination, where E$_4$ is the energy collected by
the central crystal and by its three nearest ones. Even if our
simulation of E$_4/$E$_{13}$ recovers the drop in signal
collection caused by the loss at the crystal borders, it is unable
to recover the signal spread in neighboring crystals. Therefore,
dealing with these distributions as parameters, effective
broadening due to geometrical effects is introduced in the
reconstructions, and good events are eliminated.
\begin{figure}[h!]
\begin{center}
\includegraphics[scale=0.7]{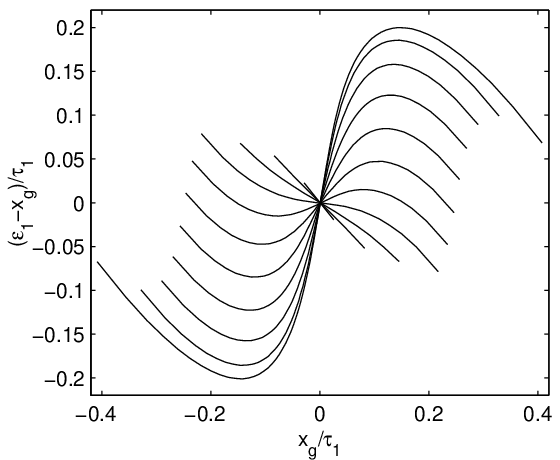}
\includegraphics[scale=0.7]{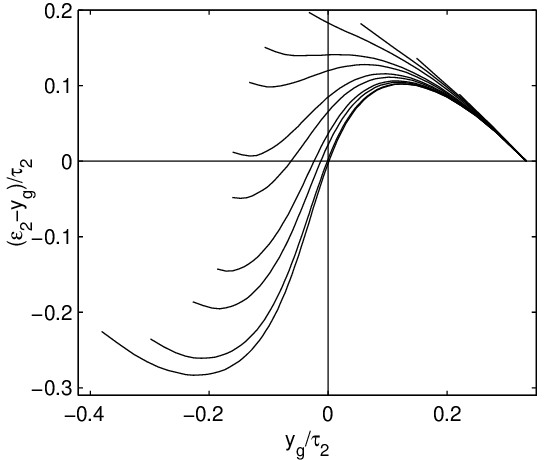}
\end{center}
\caption{\em The COG error $(\varepsilon_1-x_g)$ in function of
$x_g$ (left), each curve plots a specific $\varepsilon_2$-value.
The COG error $\varepsilon_2-y_g$ in function of $y_g$ (right),
each curve plots a specific $\varepsilon_1$-value. }
\label{fig:diciannove}
\end{figure}

The COG error in CB, as shown in Fig.~\ref{fig:diciannove}, is
very large, and without correction, errors of the order of
$\tau_1/4$ are often encountered. This is due to the very narrow
core of our---very realistic---shower model that concentrates a
large fraction of signal in the central crystal. The plots in
Fig.~\ref{fig:diciannove} are rounded at the borders due to the
signal collected by the lateral detectors, and the range of
$x_g$-values covers almost all the central detector. In contrast
to Fig.~\ref{fig:diciassetteb}, $(\varepsilon_1-x_g)$ has a strong
dependence on $\varepsilon_1$ and $\varepsilon_2$, and, similarly
$(\varepsilon_2-y_g)$. It is evident that the correction of these
COG errors is a very complex operation, and the programs developed
by CB collaboration were accordingly complex.

This example illustrates one of the more complicated cases, but
the application of our equations is not more involved than the
case of an array of square detectors. The equations and
considerations in Section 3.3 are crucial to rapidly generate the
simulation.

\section{Conclusions}

The COG error calculation for a two-dimensional array of sensors
has been explored in detail. All the relevant equations have been
discussed to allow safe applications with similar or different
systems. This approach generalizes the results of~\cite{landi01},
with the evident complexity of the two-dimensional geometry. All
the equations contain double infinite sums and look very involved.
The sums converge rapidly and a fairly limited number of terms are
required to get satisfactory results; the applications and figures
presented were obtained with only a few lines of
MATLAB~\cite{matlab} code.

Forms of crosstalk which save the signal collection efficiency,
and those in which the COG algorithm is free of discretization
errors for any form of signal distributions may be isolated. The
set of special forms is now much larger than in the
one-dimensional case. These properties are collected in easily
verifiable conditions on the function g$(\mathbf{x})$.

The set of properties of the g$(\mathbf{x})$-functions without
crosstalk made it possible to prove that orthogonal and complete
bases can be extracted from their Fourier transforms, giving a
nonseparable two-dimensional form of the
Wittaker-Kotel'nikov-Shannon theorem.

A type of crosstalk, which turned out to be largely independent of
the shape and periodicity of the array, was detected. This
crosstalk only generates an approximate cancellation of the COG
discretization error for any signal distribution. The effect is
given by the first zero of the J$_1$-Bessel function which
suppresses the first element of the COG error series. The fast
decrease with the frequency of the Fourier-Series terms forming
the COG error assigns an important contribution to the first term.
The remaining terms contribute only slightly to the total
COG-error. Not being exact, the result must be tested for the
specific problem, but from the structure of the terms forming the
COG errors, it can be inferred that the error reduction is
substantial for any signal distribution.

To prove the previous general (or approximate) results, we worked
with an infinite array of identical detectors, these equations can
be used for detector optimizations.

In the experiments, reconstructions with a selection of detectors
is a standard procedure to reduce the noise. To study the effects
of these selections, we developed methods to handle a finite set
of detectors. It is evident that the exact results are only of
approximate validity in these cases. If the crosstalk completely
suppresses the COG discretization error for any signal
distribution in an infinite array, the suppression could not be
complete for a finite set of detectors and could be (slightly)
dependent upon the signal distribution.

The simulation of the energy reconstruction in an em-calorimeter
such that of the CMS experiment highlights a position dependence
of the reconstruction efficiency. This dependence does not appear
to be recoverable using the standard tools for extracting
parameters from ratios of signal collected by a different number
of detectors. The $\pmb\varepsilon$-dependence of these
ratios does not allow their use as parameters. The fluctuations of
the Monte Carlo simulations cover the distribution of their mean
values due to the shower position.
%



\begin{thebibliography}{99}
\bibitem{landi01} G. Landi, Nucl. Instr. and Meth. A 485 (2002)
698 (also arXiv:1908.04447).
\bibitem{crystball} E.D. Bloom and C.W. Peck, Annu.
Rev. Nucl. Part. Sci. 33 (1983) 143.
\bibitem{libroFT} R.N. Bracewell, {\em The Fourier Transform and Its
Application}. (McGraw-Hill, New York, NY, 1986).
\bibitem{Escher}M.C. Escher, J.W. Vermeulen, {\em Escher on Escher: Exploring the Infinite.}
 (H.N.Adams. New York, NY, 1989).
\bibitem{Penrose}R. Penrose, {\em The Emperor's New Mind: Concerning Computers,
Minds, and the Laws of Physics}. (Oxford University Press, Oxford,
UK, 1989)
\bibitem{libroFT2} D.C. Champeney, {\em A Handbook of Fourier
Theorems}. (Cambridge University Press, Cambridge, UK, 1987).
\bibitem{DEP} B.V. Delft Electronische Producten, Roden, The
Netherlands.
\bibitem{mathematica} MATHEMATICA 3. Wolfram Research, Inc.
\bibitem{bookFT2} H. Dym, H.P. McKean, {\em Fourier Series and
Integrals.} (Academic Press, London, 1972).
\bibitem{WKS} A.J. Jerri, Proceedings IEEE 65, 1565 (1977).
\newline
J.M. Whittaker, Proc. Math. Soc. 1, 169 (1929).
\newline
V.A. Kotel'nikov, Izd. Red. Upr. Svyazi RKKA (Moscow, URSS, 1933).
\newline
C.E. Shannon, Proc. IRE 37, 10 (1949).
\bibitem{mallat} S. Mallat, {\em A Wavelet Tour of Signal
Processing.} (Academic Press, London, UK, 1999).
\bibitem{CMS1} CMS. {\em The Tracker Project}. TDR. CERN/LHCC 98-3.
\bibitem{CMS2} CMS. {\em The Electromagnetic Calorimeter}. TDR.
CERN/LHCC 97-33.
\bibitem{lednev} A.A. Lednev, Nucl. Instr. Meth. A 366, 298 (1995).
\bibitem{klanner} S. Hillert et. al., Nucl. Instr. Meth. A 458,
710 (2001).
\bibitem{particledatabook} E. Longo and I. Sestili, Nucl. Instr.
Meth. 128, 283 (1975).
\bibitem{CBBNL} T.D.S. Stanislaus et al., Nucl. Instr. Meth. A 462,
 463 (2001).
\bibitem{CBmio} D. Antreasyan et al., Z. Phys. C 48, 553 (1990).
\bibitem{matlab} MATLAB 6.1, The MathWorks, Inc.
\end{thebibliography}
\end{document}